\documentclass[sigconf,letterpaper]{acmart}

\usepackage[T1]{fontenc}
\usepackage[acronym,nowarn]{glossaries}
\usepackage[]{todonotes}
\usepackage{url}
\usepackage{marvosym}
\usepackage{ragged2e}
\usepackage{array}
\usepackage{booktabs}
\usepackage{subcaption}
\usepackage{newfloat}
\usepackage{footmisc}
\usepackage{soul}

\graphicspath{{img/}}

\newcolumntype{L}[1]{>{\raggedright\arraybackslash}p{#1}}
\newcolumntype{R}[1]{>{\raggedleft\arraybackslash}p{#1}}

\newacronym{as}{AS}{Autonomous System}
\newacronym{bgp}{BGP}{Border Gateway Protocol}
\newacronym{bod}{BOD}{Binding Operational Directive}
\newacronym{cc}{C\&C}{Command and Control}
\newacronym{cdf}{CDF}{Cumulative Distribution Function}
\newacronym{dbl}{DBL}{Domain Black List}
\newacronym{ddos}{DDoS}{Distributed Denial of Service}
\newacronym{dga}{DGA}{Domain Generation Algorithm}
\newacronym{dhcp}{DHCP}{Dynamic Host Configuration Protocol}
\newacronym{dhs}{DHS}{Department of Homeland Security}
\newacronym{dkim}{DKIM}{DomainKeys Identified Mail}
\newacronym{dmarc}{DMARC}{Domain-based Message Authentication, Reporting \& Conformance}
\newacronym{dnsbl}{DNSBL}{DNS based Blackhole List}
\newacronym{dns}{DNS}{Domain Name System}
\newacronym{dos}{DoS}{Denial of Service}
\newacronym{fp}{FP}{False Positives}
\newacronym{hsts}{HSTS}{HTTP Strict Transport Security}
\newacronym{ids}{IDS}{Intrusion Detection System}
\newacronym{ipam}{IPAM}{IP Address Management}
\newacronym{ip}{IP}{Internet Protocol}
\newacronym{isp}{ISP}{Internet Service Provider}
\newacronym{kisa}{KISA}{Korean Information Security Agency}
\newacronym{ml}{ML}{Machine Learning}
\newacronym{mta}{MTA}{Mail Transfer Agent}
\newacronym{ocr}{OCR}{Optical Character Recognition}
\newacronym{pdns}{pDNS}{passive DNS}
\newacronym{ptr}{PTR}{Pointer Record}
\newacronym{rbl}{RBL}{Real-time Blackhole List}
\newacronym{rdns}{rDNS}{Reverse DNS}
\newacronym{slaac}{SLAAC}{StateLess Address Auto Configuration}
\newacronym{sld}{SLD}{Second Level Domain}
\newacronym{smtp}{SMTP}{Simple Mail Transfer Protocol}
\newacronym{spf}{SPF}{Sender Policy Framework}
\newacronym{sql}{SQL}{Structured Query Language}
\newacronym{ssl}{SSL}{Secure Sockets Layer}
\newacronym{svm}{SVM}{Support Vector Machine}
\newacronym{tld}{TLD}{Top Level Domain}
\newacronym{tls}{TLS}{Transport Layer Security}
\newacronym{tp}{TP}{True Positives}
\newacronym{ttl}{TTL}{Time To Live}
\newacronym{url}{URL}{Uniform Resource Locator}
\newacronym{ut}{UT}{University of Twente}

\makeglossaries{}
\glsdisablehyper{}

\DeclareFloatingEnvironment[fileext=sam, name=Example, placement=htb, within=none]{example}

\iffalse
    \newcommand\reviewfix[1]{{\color{green}\sffamily\bfseries [RF: #1]}}
\else
    \newcommand\reviewfix[1]{}
\fi

\keywords{Reverse DNS, DHCP, Privacy, Tracking}

\acmYear{2022}
\copyrightyear{2022}
\setcopyright{rightsretained}
\acmConference[IMC '22]{ACM Internet Measurement Conference}{October 25--27, 2022}{Nice, France}
\acmBooktitle{ACM Internet Measurement Conference (IMC '22), October 25--27, 2022, Nice, France}
\acmPrice{}
\acmDOI{10.1145/3517745.3561424}
\acmISBN{978-1-4503-9259-4/22/10}

\ccsdesc[500]{Networks~Naming and addressing}
\ccsdesc[500]{Networks~Network management}
\ccsdesc[500]{Networks~Network privacy and anonymity}
\ccsdesc[100]{Networks~Network dynamics}

\title[Saving Brian's Privacy: the Perils of Privacy Exposure through Reverse DNS]{Saving Brian's Privacy:\\ the Perils of Privacy Exposure through Reverse DNS}

\author{Olivier van der Toorn} 
\affiliation{%
  \institution{University of Twente} 
  \country{}}
\email{o.i.vandertoorn@utwente.nl}

\author{Roland van Rijswijk-Deij}
\affiliation{%
  \institution{University of Twente} 
  \country{}}
\email{r.m.vanrijswijk@utwente.nl}

\author{Raffaele Sommese}
\affiliation{%
  \institution{University of Twente} 
  \country{}}
\email{r.sommese@utwente.nl}

\author{Anna Sperotto}
\affiliation{%
  \institution{University of Twente} 
  \country{}}
\email{a.sperotto@utwente.nl}

\author{Mattijs Jonker}
\affiliation{%
  \institution{University of Twente} 
  \country{}}
\email{m.jonker@utwente.nl}

\begin{document}


\begin{abstract}
Given the importance of privacy, many Internet protocols are nowadays designed
with privacy in mind (e.g., using TLS for \emph{confidentiality}). Foreseeing all
privacy issues at the time of protocol design is, however, challenging and may become near
impossible when interaction out of protocol bounds occurs.
One demonstrably not well understood interaction occurs when DHCP
exchanges are accompanied by automated changes to the global DNS (e.g., to
dynamically add hostnames for allocated IP addresses).
As we will substantiate, this is a privacy risk:
one may be able to infer device presence and network dynamics from
virtually \emph{anywhere} on the Internet --- and even identify and track individuals --- even if other mechanisms to limit
tracking by outsiders (e.g., blocking pings) are in place. 

We present a first of its kind study into this risk. We
identify networks that expose client identifiers in reverse DNS records and study the relation between
the presence of clients and said records. Our results show a strong
link: in 9 out of 10 cases, records linger for at most an hour, for a selection of academic,
enterprise and ISP networks alike.
We also demonstrate how client patterns and network dynamics can be learned, by
tracking devices owned by persons named \emph{Brian} over time, revealing
%
shifts in work patterns caused by COVID-19 related work-from-home measures, and by
determining a good time to stage a heist.

\end{abstract}

    \maketitle{}

\section{Introduction}
\label{sec:introduction}

Privacy violations frequently headline the news. Notorious incidents from
the past years have often involved threat actors, questionable data processing
practices, or human error.
Given the importance of privacy, many Internet protocols are nowadays designed with
it in mind (e.g., using TLS for \emph{confidentiality}). Still, it is
challenging to foresee all privacy issues at protocol design, and this may be
infeasible if interaction out of protocol bounds can occur.

One protocol that has prompted privacy concerns is the \gls{dhcp}, which is a
network management protocol that is used to dynamically assign IP addresses
to devices on a network. \gls{dhcp} uses a client-server model and allows for client devices
to send optional communication parameters to the server.
A number of research efforts have focused on DHCP privacy, demonstrating that
locally monitoring and sniffing DHCP messages -- which can contain unique client
identifiers -- enables geotemporal tracking of clients, even if clients move
between networks~\cite{tront2011,bernardos2015,aura2007}.  Groat et
al.~\cite{groat2011} even introduce the idea of remotely monitoring \gls{dhcp} client devices, using
compromised DHCPv6 relays inside the monitored network.

\gls{dhcp} exchanges can be accompanied by changes to the \gls{dns}.
\gls{dhcp} servers can update \emph{local} name
services, to associate devices on the local network with a name. Servers can also make changes to the \emph{global}
\gls{dns}, for example by adding hostnames (i.e., Pointer \texttt{(PTR)}
records) for allocated IP addresses. \texttt{PTR} records can then be publicly accessed using \gls{rdns} lookups, mapping IP addresses to hostnames.
RFC 7844~\cite{RFC7844} recognizes the privacy risk of carrying over unique client
identifiers from \gls{dhcp} options to \gls{dns}, but the extent to which this
happens in practice has received little attention in the research literature.
Furthermore, making automated changes to \gls{dns} records based
on \gls{dhcp} exchanges is in itself a privacy
risk that seems to have stayed under the radar so far. As we will demonstrate, this
practice allows network dynamics to be observed from virtually \emph{anywhere}
on the Internet.
\reviewfix{e-sum-5}The risks resulting from this are that one can track the presence of
client devices in networks, may be able to reliably identify specific devices and tie
these to persons, and might even be able to track clients across multiple
networks.

In this paper, we perform a first of its kind study into these risks.
\reviewfix{e-sum-2}Our goal is to substantiate that the interaction between
DHCP and DNS leads to unwanted leaks of potentially privacy-sensitive
information, which is then disclosed to the public Internet due to the open
nature of the DNS. We demonstrate the existence of this threat by investigating
\gls{rdns} data. \reviewfix{e-sum-1}Our results show that there is a disparity between the
guidance in standards with respect to privacy and \gls{dhcp} and \gls{dns}
interaction, versus the actual implementation in practice. This disparity is
what allows for sensitive information to leak to the public Internet.

The contributions of this paper are:
\begin{enumerate}
    \item We show that \gls{dns} records contain unique identifiers in practice,
	  even including sensitive information such as client device types and device owner names.
    \item We demonstrate that networks of varying types (academic, enterprise, ISP) expose such information.
    \item We analyze the relation between the presence of dynamically added hostnames in the \gls{dns} and the presence of client devices on networks.
    \item We demonstrate that outsiders can use reverse \gls{dns} to track specific clients and learn network dynamics.
    \item We discuss possible causes and ways to mitigate risk.
\end{enumerate}

The remainder of this paper is organized as follows. In
Section~\ref{sec:related_work} we provide background information and discuss
related work. We introduce our data sets in Section~\ref{sec:datasets}.
We detail how we identify networks that expose dynamic behavior
Section~\ref{sec:dynamicity}.  In Section~\ref{sec:identifying_leaks} we focus
on leaking privacy-sensitive information in \gls{rdns} records.  We look at
timing of dynamically added \gls{rdns} entries in
Section~\ref{sec:timeliness}.
Then, in Section~\ref{sec:case_studies}, we present a number of case studies to show how
client patterns and network dynamics can be learned.  In
Section~\ref{sec:discuss} we discuss our findings and possible steps toward
mitigating the privacy risk. Finally, we document ethical considerations in
Section~\ref{sec:ethical_considerations} and conclude in Section~\ref{sec:conclusion}.


\section{Background \& Related Work}
\label{sec:related_work}

\subsection{Background}
\label{ssec:related_work:background}
In this section we provide background information on \emph{reverse} and
\emph{forward} \gls{dns}, the \gls{dhcp}, and \gls{ipam} systems.

\paragraph{DNS}

The \gls{dns} is a critical component of the Internet and can be seen
as its \textit{phonebook}. It is responsible for
translating between human-readable names and IP addresses.
The \gls{dns} is operated as a distributed hierarchical
database, in which parts of the namespace are delegated to different
parties~\cite{rfc1034}.
The \gls{dns} is typically used to translate domain names to IP addresses,
which is referred to as \emph{forward} DNS. The \gls{dns} also
enables \emph{reverse} resolution, in which an IP address is translated to a hostname.
The \gls{dns} uses special zones for the purpose of the latter. The \texttt{in-addr.arpa.}
zone is used to translate IPv4 addresses to hostnames~\cite{rfc1035}.
This zone contains so-called \texttt{PTR} (Pointer) records, which one can query
for by using the reversed form of the IP address one wishes to translate.
With the reversed form, the \gls{dns} can be queried similarly as with forward
\gls{dns}, except that a \texttt{PTR} record is requested, rather than an
\texttt{A} record.
Example~\ref{example:related_work:background:revdns} provides an example of an
IP address and its reversed form for a \texttt{PTR} query.

\begin{example}
    \centering
    \resizebox{.8\columnwidth}{!}{%
    \begin{tabular}{l c}
        What? & Value \\
        \toprule
        IP address to translate   & \texttt{93.184.216.34} \\
        Reversed \gls{dns} query  & \texttt{34.216.184.93.in-addr.arpa.} \\
        \bottomrule
    \end{tabular}
    } 
    \caption{Reverse IPv4 Example}
    \label{example:related_work:background:revdns}
\end{example}

\paragraph{DHCP}
The \glsfirst{dhcp} is a network management protocol that can be used to dynamically
assign IP addresses to devices on a network, using a client-server model. When
a client device wishes to join the network using \gls{dhcp}, it issues an address
request to \gls{dhcp} servers, either via broadcast or unicast in the event a server
address is already known.
The \gls{dhcp} server can offer and subsequently allocate an IP address
to the client for a set amount of time. Upon allocation, the client is told
which address it is allowed to use and for how long. This allocation is called the
\gls{dhcp} lease and before it expires, the client can request
renewal. If a lease expires, the associated IP address is considered
reallocable.
When a client leaves the network it can signal to the \gls{dhcp} server,
through a so-called release message, that it is in the process of leaving the network.
The server can then reallocate the IP address sooner.  Release messages are not
always sent, as clients can go out of range, or users can unplug devices from
the network abruptly.

\gls{dhcp} client-server exchanges involve more information than the allocated
IP address and expiration time. The server can convey communication
parameters such as the default gateway. The client in turn can send
information such as an optional \texttt{Host Name}~\cite{RFC2132} or
\texttt{Client FQDN}~\cite{RFC4702}.
The prior is commonly used by \gls{dhcp} servers to identify hosts and also to update
the address of the host in \emph{local} name services. The latter would allow the
server to update \emph{global} name services, if the client so desires (cf. $\S$3.3 in~\cite{RFC4702}). 

\paragraph{IPAM Systems}
\glsfirst{ipam} systems allow network operators to manage various
parts of IP address infrastructure. These systems are typically used in larger
enterprises, where manually managing IP address space (e.g., assigning subnets
or IP addresses) is no longer feasible. \gls{ipam} systems can be used to manage
\gls{dhcp} as well \gls{dns}.

\paragraph{Interplay between DHCP and DNS}
\gls{dhcp} and \gls{dns} can be linked together, through \gls{ipam} systems or by
other means. If they are linked, then when a client requests a \gls{dhcp} lease and is
allocated an IP address, various changes to the \gls{dns} related to the IP
address are made automatically.
For example, the \gls{dhcp} server can update information for the
allocated IP address in local name services, which the server may do on the basis of the
previously mentioned, client-provided \texttt{Host Name}.\footnote{Note that
\texttt{hostname} is commonly used to refer to the name to which
an IP address translates (i.e., the name in its \texttt{PTR} record). To avoid
ambiguity, we use \texttt{Host Name} to refer to the \gls{dhcp} client
parameter.}

Automated changes to \gls{dns} are not limited to \emph{local} name services per se.
Changes to the \emph{global} \gls{dns} can also be made. For example, upon allocation of an IP address, a
hostname can be associated with said address by adding a \texttt{PTR}
record to the global \gls{dns}.
Evidently, if changes to the (public) \gls{dns} are made as client devices join or leave a network, one
may be able to infer network dynamics by capturing \gls{dns} changes.
The privacy risk further increases if client-provided parameters are used in
\gls{dns} updates, as these may allow for specific devices to be identified and their presence
or absence to be tracked.
RFC 7844~\cite{RFC7844} recognizes the potential risk of carrying-over unique
identifiers to the DNS. As we will show, the makes, models and even owner names
of devices running the \gls{dhcp} client can appear in \gls{dns} data (e.g.,
\emph{Brian's iPhone}).

\subsection{Related Work}
\label{ssec:related_work:related_work}

We consider several types of related work: gaining information from
hostnames in \gls{rdns}, \gls{dns} and privacy, and \gls{dhcp} and privacy.

\paragraph{Information in hostnames}
It is known in the literature that reverse \gls{dns} can reveal information about hosts.
Various works exist in which authors successfully use hostnames
to gain insights into Internet infrastructure and topology. Central to
such works is the notion that hostnames can encode meaningful information.
Chabarek et al.~\cite{chabarek2013} used \gls{rdns} data to study part
of the Internet core, by inferring link speeds of router and switch interfaces
from hostnames.
Huffaker et al.\ instead extract geographic information, using a dictionary of city names and
airport codes~\cite{huffaker2014drop}. Follow-up works by Luckie et al. focus on learning how to extract autonomous systems and
network names from hostnames~\cite{luckie2021-conext,luckie2021-aintec}.

The aforementioned works focus mostly on core infrastructure such as routers. Lee et
al.~\cite{lee2017} instead shift focus to end-users, i.e.,
customers of \glspl{isp}. In their paper, they study means to infer the
connection types of hosts in access networks.
Zhang et al.~\cite{zhang2021inferring}, in turn, infer and geolocate topology in
regional access networks, with the aim of studying architectural choices made
by ISPs.

While all these works extract meaningful information from
hostnames, to the best of our knowledge no works exist that consider what
can be learned about networks by observing \emph{automated} and \emph{continual} changes to
\gls{rdns} records. We bridge this gap and reveal that \gls{rdns} changes can provide
insight into network dynamics and client behaviors.
A common assumption in related work also seems to be that meaningful information
is included in \gls{rdns} by network operators on purpose. We instead
substantiate that reverse \gls{dns} can inadvertently contain privacy-sensitive
information.

\paragraph{DNS and Privacy}
The \gls{dns} is no stranger when it comes to privacy considerations on the
Internet. In fact, \gls{dns} privacy has received a lot of attention over the past
years. Often, the focus is on keeping interactions with the \gls{dns} confidential, or on
aspects of \gls{dns} data sharing and processing~\cite{imana2021,
rfc9076,kang2016,rijswijk-deij2019}.
An example is QNAME-minimization~\cite{rfc7816,vries2019}, which helps improve
privacy by minimizing information sent from the recursive to upstream name
servers. Other examples involve efforts to add confidentiality by
encrypting queries and answers, for example via DNS-over-HTTPS (DoH) or
-over-TLS (DoT).
Generally speaking, studies that relate to the \gls{dns} often involve
\emph{forward} \gls{dns}. The \emph{reverse} side of \gls{dns} is less
frequently studied.
Tatang et al.~\cite{tatang2019large} studied privacy leaks in \gls{rdns} to a
certain extent, after observing that some misconfigured name servers
provide outsiders with answers to \texttt{PTR} queries for private IP addresses (e.g., RFC
1918~\cite{rfc1918}), if such addresses are used inside the
networks of the misconfigured servers. In their paper they characterized the country and network
distribution of such servers, and studied privacy-sensitive patterns in hostnames, revealing end-user devices
as well as security-critical infrastructure such as firewalls.

\paragraph{DHCP and Privacy}
The \gls{dhcp} protocol has given rise to privacy concerns, which
led to discussions in the Dynamic Host Configuration (DHC) working group. These discussions
resulted in RFC 7844~\cite{RFC7844}, which recognizes that \gls{dhcp} client
messages can contain unique client identifiers. Such identifiers can be used to track
clients, even if devices take care of randomizing other link-layer identifiers
such as MAC addresses. RFC 7844 also recognizes that identifiers
can carry-over to the DNS. They propose anonymity profiles, which
minimize disclosure of client-identifying information in \gls{dhcp} messages.

Tront et al.~\cite{tront2011} proposed using a dynamic DHCP unique identifier (DUID)
based on the same randomization technique used in IPv6 privacy extensions.
Groat et al.~\cite{groat2011} show
how the use of DHCPv6 to overcome the privacy issue of SLAAC deployment
can still lead to the possibility to track users because of the use of
static DUIDs. They also note that remote tracking may be possible, via compromised
DHCPv6 relays that forward messages to attackers.
Bernardos et al.~\cite{bernardos2015} showed how randomization of L2
addresses was a convenient solution to mitigate location privacy issues on
public Wi-Fi connections, evaluating also user experience and potential IP address
pool exhaustion.
Aura et al.~\cite{aura2007} investigated how DHCP can be used to
provide mobile clients with authenticated network location information, which
clients can then use to decide how to behave in specific networks. In their
paper, they consider the privacy of mobile users, by minimizing client
information in \gls{dhcp} messages, at least until the network has been
authenticated.
Finally, while the previously discussed work by Tatang et
al.~\cite{tatang2019large} did find patterns in the DNS that likely resulted
from DHCP carry-over, the authors appear to not have considered the role of
DHCP.

The literature has thus established that \gls{dhcp} messages create opportunities
to track client devices, even between subnets and networks. Central to most of
these works is the ability to observe messages, which requires observer
presence in network.  Our work instead focuses on the risk associated with the interplay
between \gls{dhcp} and \gls{dns}. We extend and substantiate the risk that
theoretically existed and study the problem in the wild.

\section{Data sets}
\label{sec:datasets}

We use three data sets in this paper: two large-scale data sets of IPv4
\gls{rdns} measurements, and a smaller data set with ICMP and \gls{rdns}
measurements that we collect ourselves.

\paragraph{Full address-space reverse DNS measurements}
The bulk of our analysis relies on measurement data from reverse \gls{dns}
measurements that cover the entire IPv4 address space. Several projects make
\gls{rdns} data sets available for research. In this study, we use data sets
collected by Rapid7's Project Sonar~\cite{rapid72021} and by the OpenINTEL
project~\cite{openintel2021}. The Rapid7 data set is collected on a single
weekday every week and OpenINTEL collects daily snapshots. Given that our goal
is to show evidence of dynamic address assignments relating to human behavior,
we prefer the data from OpenINTEL because of its daily measurement
frequency. Where we need data that predates the first collection date
obtainable from OpenINTEL, we use data from Rapid7 instead. Jointly, both data sets
cover the period of our study: 2019-10-01 to 2021-12-31.
Table~\ref{tbl:datasets:oi_rapid7} shows summary statistics on the data used from Rapid7 and
OpenINTEL. The table details the number of data points in each data set as
well as the daily average number of unique {\tt PTR} records observed by each
measurement.

\paragraph{Reactive fine-grained measurement}

The third data set that we use in this work is a custom, supplemental
measurement, described in Section~\ref{ssec:methodology:active_measurement}.
This data set includes both data from a dedicated ICMP measurement and
fine-grained data from a reactive \gls{rdns} measurement. The supplemental
measurement was performed from 2021-10-27 to 2021-11-16. Summary statistics for
this supplementary data set are given in Table~\ref{tbl:datasets:active_measurement}.

\begin{table}
    \caption{Statistics for the data sets that we obtained from Rapid 7 and OpenINTEL.}
    \label{tbl:datasets:oi_rapid7}
    \begin{center}
        \resizebox{\columnwidth}{!}{%
            \begin{tabular}{l c c r r}
\toprule
{}           &  Start date &    End date &  Total \# responses &  \# unique PTRs \\
\midrule
Rapid7 Sonar &  2019-10-01 &  2021-01-01 &      77\,G &  1,381\,M \\
OpenINTEL    &  2020-02-17 &  2021-12-01 &     396\,G &  1,356\,M \\
\bottomrule
\end{tabular}

        }
    \end{center}
\end{table}

    \section{Identifying Dynamicity Exposure}
\label{sec:dynamicity}
\reviewfix{e-sum-6 - restructuring}

In this section, we discuss how we identify networks that expose dynamic behavior through \gls{rdns} entries. We then proceed to apply our identifying methodology to our data set.

\subsection{Methodology}
\label{ssec:methodology:network_identification}

To study whether networks dynamically add and expose privacy-sensitive
information through \gls{rdns}, we first need to identify which networks
exhibit signs of dynamic behavior in our data sets. 
\reviewfix{e-sum-3-2}We focus only on such networks because we aim at investigating temporal patterns of client devices.
Other networks can still expose privacy-sensitive information in \texttt{PTR}
records (i.e., hostnames) in a non-dynamic sense.  However, these networks
cannot be leveraged to learn temporal patterns related to clients, therefore we
do not include them in our analysis.
To identify dynamic networks, we use the daily data sets obtained from
OpenINTEL and apply a set of heuristics in three steps detailed below.

\vspace{0.3em}
\noindent
{\bf Step 1:} First, we perform a daily analysis over data covering a
three-month period. We group results by {\tt /24} prefix and compute the
unique number of IP addresses for which we see a {\tt PTR} record on each
day. We then discard the {\tt /24} prefixes for which we never observe more
than 10 addresses a day over the three-month period under consideration. For the prefixes
for which we do, we also record the maximum number of daily IP addresses per
{\tt /24} over the three-month period.

\vspace{0.3em}
\noindent
{\bf Step 2:} Next, we perform a day-by-day comparison for each {\tt /24} considered in
over the three-month period and record the absolute
difference in number of IP addresses for which we observe a {\tt PTR} record.
We then compute the ``change percentage'' by dividing this
difference by the maximum number of addresses recorded in the previous step.

\vspace{0.3em}
\noindent
{\bf Step 3:} Finally, we label {\tt /24} prefixes as {\it dynamic} if the
change percentage exceeds $X\%$ on at least $Y$ days over the entire
three-month period. We set $X$ to 10 (which sets the threshold at a single
address changed for blocks with 10 or more addresses), and $Y$ to 7, based on
experiments.

\vspace{0.3em}
\noindent
{\bf Validation:} We validate our heuristic approach against our own campus
network. The addresses for this network come from a single {\tt /16} prefix
with a numbering plan in which some subprefixes are used for dynamic
allocations whereas other subprefixes contain static allocations.
We run our
approach to identify change-sensitive {\tt /24} blocks. Our method marks 40
prefixes as dynamic, and 206 prefixes as static. We confirmed these results
with our campus ICT department. The 40 prefixes we identify as dynamic are
confirmed as true positives. In addition to this, our IT department indicated
a further 83 prefixes use dynamic address assignments (\gls{dhcp}), but with static
\gls{rdns} entries (i.e., fixed-form {\tt PTR} records such as {\tt
host1234.dynamic.institute.edu}). This confirms that our heuristic approach correctly
identifies prefixes with dynamically updated {\tt PTR} records.

\vspace{0.25em}
\noindent
\reviewfix{e-sum-4, c-1}{\bf Threshold and dynamicity:} our dynamicity
methodology strongly depends on setting thresholds (for the change percentage
\textit{X} and the number of days \textit{Y}). Given the values that we chose, we discard a large
number of {\tt /24}. Our rationale behind choosing such strict thresholds is that we
want to identify dynamic networks with high confidence. As the preceding
validation involving ground truth demonstrates, our
threshold choices are reasonable. Thus, using these thresholds we establish a
lower bound of dynamicity exposing networks.

\subsection{Identifying Networks}

We start out by identifying networks that exhibit dynamic behavior using the
approach detailed in Section~\ref{ssec:methodology:network_identification}. We
use the three-month period from 2021-01 to 2021-03 to identify such networks.
Over this period, we see {\tt PTR} records for a total of 6,151,219 unique
{\tt /24} networks. Out of these, 134,451 are marked as dynamic using our
heuristic approach. \reviewfix{e-2} This result demonstrates that there is alarming evidence
of networks exposing dynamics in (global) \gls{rdns}.

\begin{figure}[t]
	\centering
	\includegraphics[width=\columnwidth]{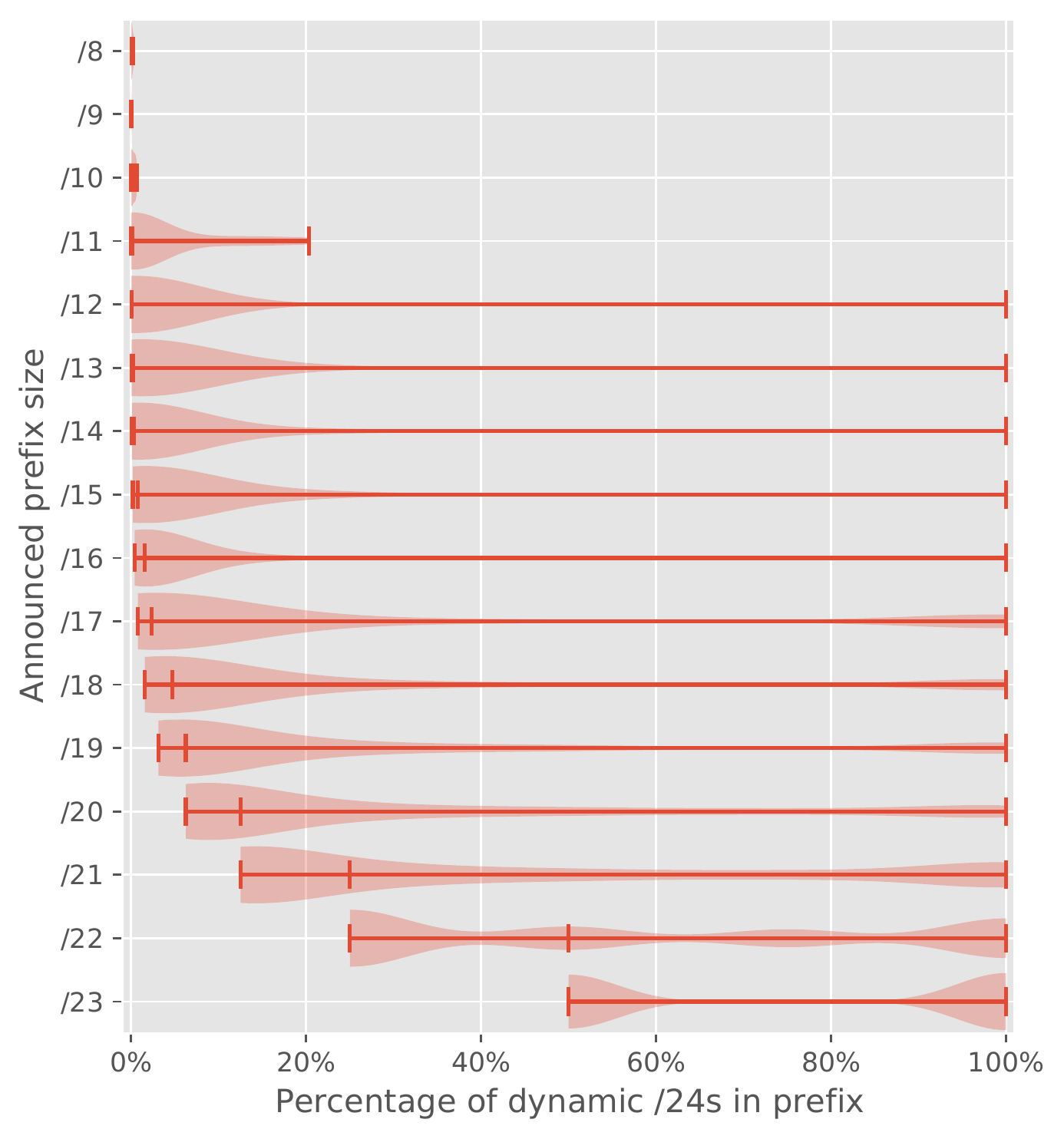}
	\caption{Distribution of the fraction of {\tt /24} prefixes that show
	dynamic \gls{rdns} behavior as part of the most-specific announced
	prefix they are part of. Ticks show the minimum, median and maximum
	number of {\tt /24} subprefixes that show dynamic behavior.}
	\label{fig:dynamic-violin}
\end{figure}

To gain a further intuition on how dynamic behavior in \gls{rdns} is visible as
part of a larger network, we map any {\tt /24} prefix that we identify as dynamic back to
the most-specific announced, covering prefix.
Figure~\ref{fig:dynamic-violin} shows the distribution of the fraction of {\tt
/24} prefixes that make up a prefix that exhibit dynamic behavior. As the plot
shows, generally speaking, only a small subset of the prefixes that make up a
network exhibit dynamic behavior. An intuition for this result is the use of
numbering plans, where specific subprefixes are used for dynamic clients
(recall how this is done in our own campus network from the validation of our
heuristic approach in Section~\ref{ssec:methodology:network_identification}).
We leave further study of external visibility of such network segmentation as
future work. Finally, we note that the result in
Figure~\ref{fig:dynamic-violin} also guides the choice of which subprefixes to
subject to our supplemental measurement, which we return to later in
Section~\ref{ssec:methodology:active_measurement}.

    \section{Identifying Privacy Leaks in Records}
\label{sec:identifying_leaks}
\reviewfix{e-sum-6 - restructuring}

In this section, we identify the publication of privacy-sensitive information
in \gls{rdns} and demonstrate an associated risk.

\subsection{Methodology}
\label{ssec:methodology:zooming_privacy}

Recall that our goal is to identify privacy-sensitive information in dynamically updated
\gls{rdns} entries. In order to zoom in on such privacy leaks, we perform
further filtering of our data sets, consisting of the following steps:

\paragraph{Extracting Common Terms}
We start by analyzing terms that commonly appear in \gls{rdns} records. To find
terms we use a regular expression that extracts words consisting of
alphabetical characters from \texttt{PTR} records, of which we can count occurrences.

\paragraph{Common Suffixes}
Hostnames for related IP addresses can have a common hostname suffix, to
which host-specific parts are prepended. Consider, e.g.,
\texttt{client1.someisp.com} and
\texttt{client2.someisp.com}.
We identify suffix keywords (\texttt{someisp} and \texttt{com} in
this example).

\paragraph{Generic Terms}
Among non-suffix keywords, we identify a number of generic terms that convey
location or router-level information. These terms are less likely to be used in client
hostname prefixes. Examples are \texttt{north} and \texttt{south}. We use these terms
to exclude router-level \texttt{PTR} records (see also~\cite{luckie2021-aintec,luckie2021-conext}).

\paragraph{Given Names}
From the remaining \texttt{PTR} records we then select those that contain
\emph{given names}, as given names can be indicative of a user client device
hostname.
The US government keeps track of and publishes names given to
newborns.\footnote{\url{https://www.ssa.gov/oact/babynames/}} We select names
for the years 2000 up to 2020, ranked by popularity over this 20-year period. We
select the top 50 most popular names.

\paragraph{Dealing with City Names}
Router-level hostnames can encode location information
such as city names~\cite{huffaker2014drop}, which can overlap with
\emph{given names} (e.g., \emph{Jackson} and \emph{Jacksonville}).
Instead of excluding such mismatches via enumeration (e.g., using a list
of city names), we count the number of unique \emph{given name} matches per
hostname suffix and require this to be above a certain threshold.  Our
reasoning here is that if dynamic client devices are present, the number of
uniquely matched \emph{given names} will be relatively larger than the
number of unique \emph{city names} in router-level hostnames.

\subsubsection{Identified Networks}
\label{ssec:methodology:identified_networks}
We apply the aforementioned steps to daily \gls{rdns} data from OpenINTEL to find networks
that likely add \gls{rdns} records for dynamic client devices and carry over
unique identifiers to the DNS. Henceforth we refer to the found networks as the 
\emph{identified networks}.

\begin{enumerate}
    \item{We start from the set of networks showing dynamic behavior, based on the heuristic approach described in Section~\ref{ssec:methodology:network_identification}.}
    \item{We then exclude \gls{rdns} entries with generic router-level terms.}
    \item{We match the remaining \texttt{PTR} records against a list of given
          names.}
    \item{We extract hostname suffixes from the results and calculate per suffix: (1) the number
          of records; (2) the number of uniquely matched given names; and (3)
          the ratio between the two.}
      \item{We select suffixes with at least 50 unique given name matches.\footnote{\label{foot:parameters}We note that our aim is not to exhaustively identify and study all dynamic networks, but rather to identify a small set which likely leak privacy-sensitive information through \gls{rdns} entries. See also Section~\ref{sec:ethical_considerations} in which we discuss ethical considerations.}}
      \item{We further require a ratio of 0.1 or more\footref{foot:parameters}, to find a variety of matched names in sizeable networks.}
\end{enumerate}

\subsection{Signs and Causes of Privacy Trouble}
\label{ssec:results:openintel}

\begin{figure*}
    \begin{center}
        \resizebox{\textwidth}{!}{%
            \includegraphics[width=\columnwidth]{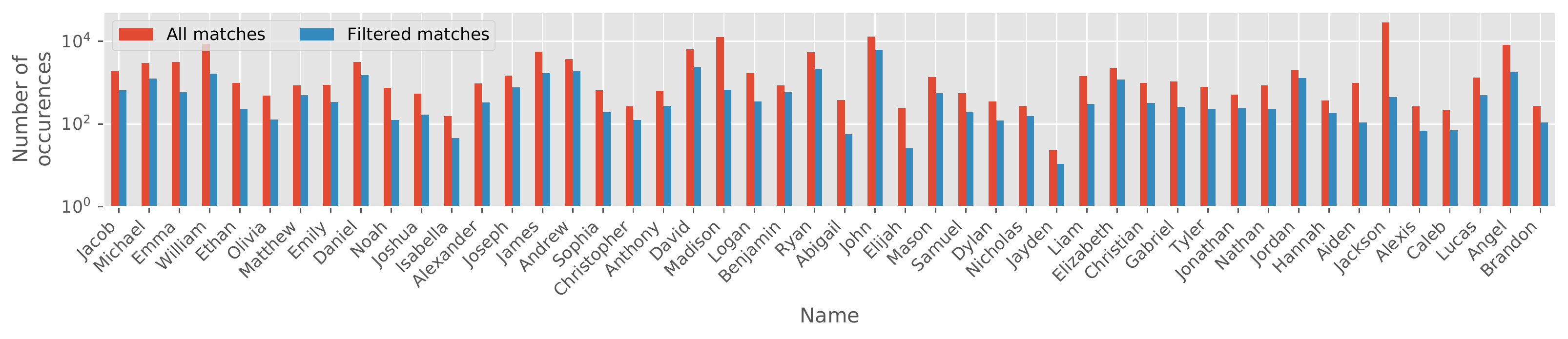}
        }
    \end{center}
    \caption{Given names for newborns (Top 50 sorted by US popularity) as observed in 
             reverse DNS entries. The plot shows the total number of matches and the number of
	     matches after filtering the networks (logarithmic scale).}
    \label{fig:results:openintel:baby-names}
\end{figure*}

We now zoom in using the additional filtering steps described in
Section~\ref{ssec:methodology:zooming_privacy}. After filtering, we identify
197 networks that meet our strict criteria. We index these networks by
hostname suffix (TLD+1) and manually classify them by network type.
We start by making a number of general observations about these networks.

\paragraph{Given names} Figure~\ref{fig:results:openintel:baby-names} shows the
number of given name matches in the \gls{rdns} data. The blue bars
account for any matching \texttt{PTR} record.
The red bars only count records that belong to networks that meet the
uncertainty-minimizing thresholds and criteria set out in
Sections~\ref{ssec:methodology:network_identification}
and~\ref{ssec:methodology:zooming_privacy}.
Figure~\ref{fig:results:openintel:baby-names} shows that given names are
generally more common in prefixes that show dynamic behavior. \reviewfix{a-7} Please note that, due to the logarithmic y-axis, the difference between the number of matches before and after filtering easily amounts to an order of magnitude.

\paragraph{Common Appearances in Hostnames}
We investigate which words commonly co-appear alongside \emph{given names} in
hostnames, postulating that these terms may enable insight into how given names
ended up in hostnames to begin with. From the common terms that occur a hundred
times or more we manually select those that we think reveal information
about client devices other than the user's \emph{given name}.

\begin{figure}
    \begin{center}
        \resizebox{\columnwidth}{!}{%
            \includegraphics[width=\columnwidth]{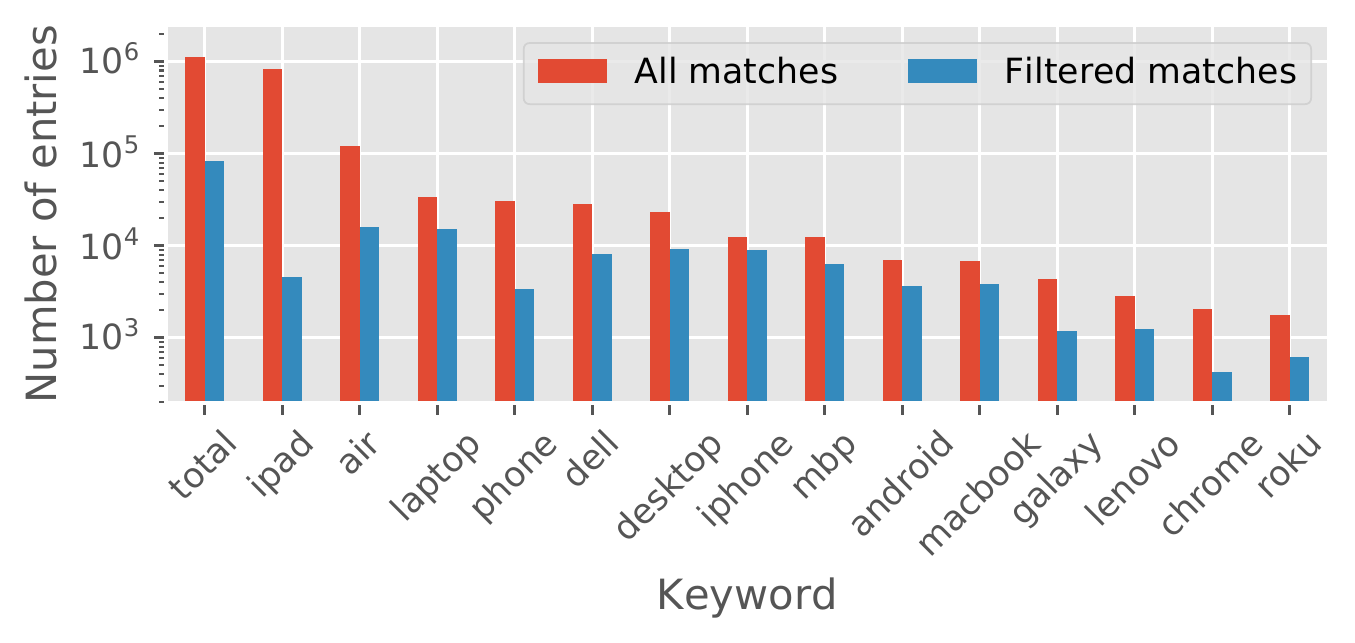}
        }
        \caption{Terms frequently in hostnames along given names,
                 before and after imposing \emph{given name} thresholds (logarithmic scale \reviewfix{a-7}).
                 The column \emph{total} is the sum, not a term.}
        \label{fig:results:openintel:keywords}
    \end{center}
\end{figure}

Figure~\ref{fig:results:openintel:keywords} shows the terms we selected, along with
the number of \texttt{PTR} records in which the terms appear, before and after imposing
the strict thresholds.
The frequent co-appearance of terms such as \texttt{iphone}, \texttt{android}
and \texttt{galaxy} are a strong indication that \gls{dhcp} clients on a
variety of mobile devices send the name of the device to the \gls{dhcp} server,
seeing as phone names can be formed of the owner's name and make or model
(e.g., \emph{Brian's iPhone}). The appearance of terms such as \texttt{laptop} and \texttt{desktop}
are indicative of behavior of \gls{dhcp} clients on other types of devices.

As previously discussed in Section~\ref{sec:related_work}, RFC 7844 recognizes the risk of
carrying over unique client identifiers from \gls{dhcp} to \gls{dns} because
these identifiers can be used to track clients~\cite{RFC7844}. Our findings do not only
demonstrate that identifiers are in fact carried over in the wild, but also
reveal that the content contained in identifiers is in itself privacy-sensitive.
For example, being able to tell the make and model of a client device may benefit
sophisticated attackers, who could use this information to pre-select relevant
exploits. Owner names, in turn, can tie IP addresses to users,
which could be used for a number of malicious purposes.

We suspect that phone and computer names are sent via the \gls{dhcp}
\texttt{Host Name}. While we do not claim that \gls{dhcp} clients are
necessarily at fault here, we do note that these terms can help identify
the makes and models of devices and that this may require mitigation steps.

\paragraph{Beyond Given Names}
While the approach we chose to identify networks hinges on the appearance of 
\emph{given names}, we recognise that some commonly co-appearing terms can also be used
independently.  As our aim is not to exhaustively identify networks, this is
not something we explore in this paper.
We note that we considered terms of three or more characters. While shorter
terms do co-appear, they add a lot of noise. As an example, consider the term
\texttt{hp}, which may indicate HP laptops and desktops, but can also be a substring
in other terms.

\paragraph{Network Types}

\begin{figure}
    \begin{center}
        \resizebox{\columnwidth}{!}{%
            \includegraphics[width=\columnwidth]{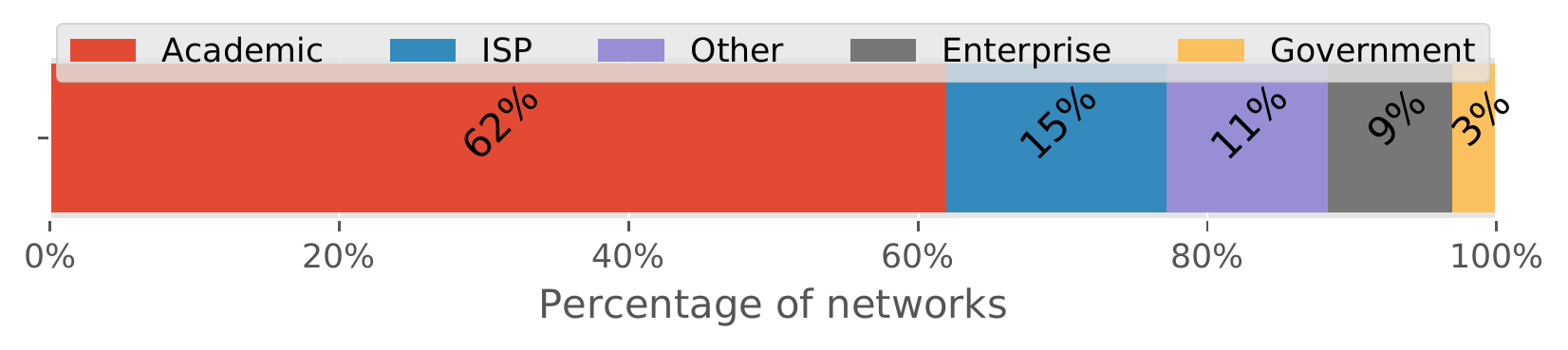}
        }
        \caption{Breakdown of the 197 networks over the types \emph{academic}, \emph{ISP}, \emph{enterprise}, \emph{government} and \emph{other}.}\label{fig:results:openintel:networktypes}
    \end{center}
\end{figure}

We use a manual selection process to infer the type of each identified network
by looking at hostname suffixes. The specific types that we identify are: \emph{academic},
\emph{ISP}, \emph{enterprise}, \emph{government} and \emph{other}.
We use regular expressions to match records
against \texttt{.edu} and \texttt{.ac}, both of which indicate \emph{academic}
use, as well as \texttt{.gov} for \emph{government} use. To find other network
types such as \emph{ISP} and \emph{enterprise} networks, we use manual
inspection. 
Figure~\ref{fig:results:openintel:networktypes} shows a breakdown of the
results for the 197 \emph{identified networks}. The majority of networks, 61.9\%, is \emph{academic} -- these contain
school, university and research institution networks. \emph{ISPs} account for
15.2\% of the networks, 9\% are \emph{enterprise} networks and 3\% are
classified as \emph{government} networks. Finally, 11.2\% have the label
\emph{other}, indicating that we were unable to clearly classify their type.         

$ $\\
\textbf{Key takeaways:} 
\emph{
There are
strong signs that networks of various types expose the presence of dynamic
clients in \gls{rdns}. A problem beyond merely carrying over unique client identifiers
from \gls{dhcp} messages to the \gls{dns} becomes apparent: the makes,
models and even owner names of client devices can be learned.}

    \section{Timing of rDNS entries}
\label{sec:timeliness}

\reviewfix{e-sum-6 - restructuring}This section first explains our supplemental
measurement to collect finer-grained timing information on dynamic client behavior
in networks that exhibit dynamicity in \gls{rdns} entries. Next, we present our
findings on the timing of \gls{rdns} records.

\subsection{Methodology}
\label{ssec:methodology:active_measurement}
The highest temporal measurement granularity of the two existing \gls{rdns}
data sets that we use is daily. This means we cannot infer sub day level
dynamicity (e.g., devices joining and leaving the network). To perform a more
detailed study of the timing of \gls{rdns} entries and attempt to capture
network dynamics beyond what can be learned from daily \gls{rdns} measurement
data, we perform a supplemental measurement against the IP address space of a
subset of \emph{identified networks}.

\paragraph{Network Selection}
\label{ssec:methodology:supplemental_selection}

To comply with the requirements from our IRB (see
Section~\ref{sec:ethical_considerations}), we select a minimal subset from the
197 \emph{identified networks} for supplemental measurement and further
validation.
We select nine networks -- three networks of the three most-common types
\emph{academic}, \emph{ISP} and \emph{enterprise} (from
Section~\ref{ssec:results:openintel}).
In the selection of three \emph{academic} networks we favor one
particular network as we have {\it a posteriori} knowledge about IP address
distribution that has utility towards one of our case studies
(Section~\ref{sec:case_studies:heist}). 

We order the list of networks of each type by
the number of \emph{given name} matches and start selecting from the top. We
perform an additional, manual inspection of \texttt{PTR} records to ensure that the networks
we select show evidence of dynamically assigned hosts.
We make a weighted selection of which address space of selected networks to target with supplemental measurement.
For large networks, we dig a little deeper to observe which IP subnet
(\texttt{/16} or more specific) contains the most dynamically assigned hosts,
and target this address space only. Whether or not networks respond to ICMP
ping scans does not factor into the selection process.

\paragraph{Measurement Mechanics}
\label{para:methodology:active_measurement:measurement}
Our supplemental measurement technique to investigate the timing of
\gls{rdns} records involves two types of probing: (1) \texttt{ICMP}
probes; and (2) finer-grained reverse \gls{dns} lookups.
We run an hourly ICMP ping scan against the selected networks to determine if
client devices have joined or left the network since the previous hour,
provided of course that the devices respond to pings.

We hypothesize that client devices on a network go through the following three phases:
\begin{enumerate}
    \item{The client joins the network and is allocated an IP address. An \gls{rdns}
	  entry is added or updated by the \gls{dhcp} server. The client device may start to respond to ping
          requests.}
    \item{The client is active on the network. In this phase it keeps responding to pings and the \texttt{PTR} remains unchanged.}
    \item{The client leaves the network and no longer responds to
          pings. The address may be deallocated and the \texttt{PTR} may be changed or removed. This is subject to
          behavior of the \gls{dhcp} server and may also depend on whether
          or not the client releases its lease (see Section~\ref{ssec:related_work:background}).}
\end{enumerate}

\begin{table}
    \caption{Reactive measurement and back off strategy.}
    \label{tbl:active_measurement:frequencies}
    \begin{center}
        \resizebox{\columnwidth}{!}{%
            \begin{tabular}{@{} L{8.5cm} @{}}
    \toprule
    Number of measurements and measurement intervals \\
    \midrule
    12 times in the 1\textsuperscript{st} hour at 5-minute intervals \\
    $\hookrightarrow$ 6 times in the 2\textsuperscript{nd} hour at 10-minute intervals \\
    \hspace{0.8em}$\hookrightarrow$ 3 times in the 3\textsuperscript{rd} hour at 20-minute intervals \\
    \hspace{0.8em}\hspace{0.8em}$\hookrightarrow$ 2 times in the 4\textsuperscript{th} hour at 30-minute intervals \\
    \hspace{0.8em}\hspace{0.8em}\hspace{0.8em}$\hookrightarrow$until client goes offline once at 60-minute intervals \\
    \bottomrule
\end{tabular}

        }
    \end{center}
\end{table}

\begin{figure}[t]
	\centering
	\includegraphics[width=\columnwidth]{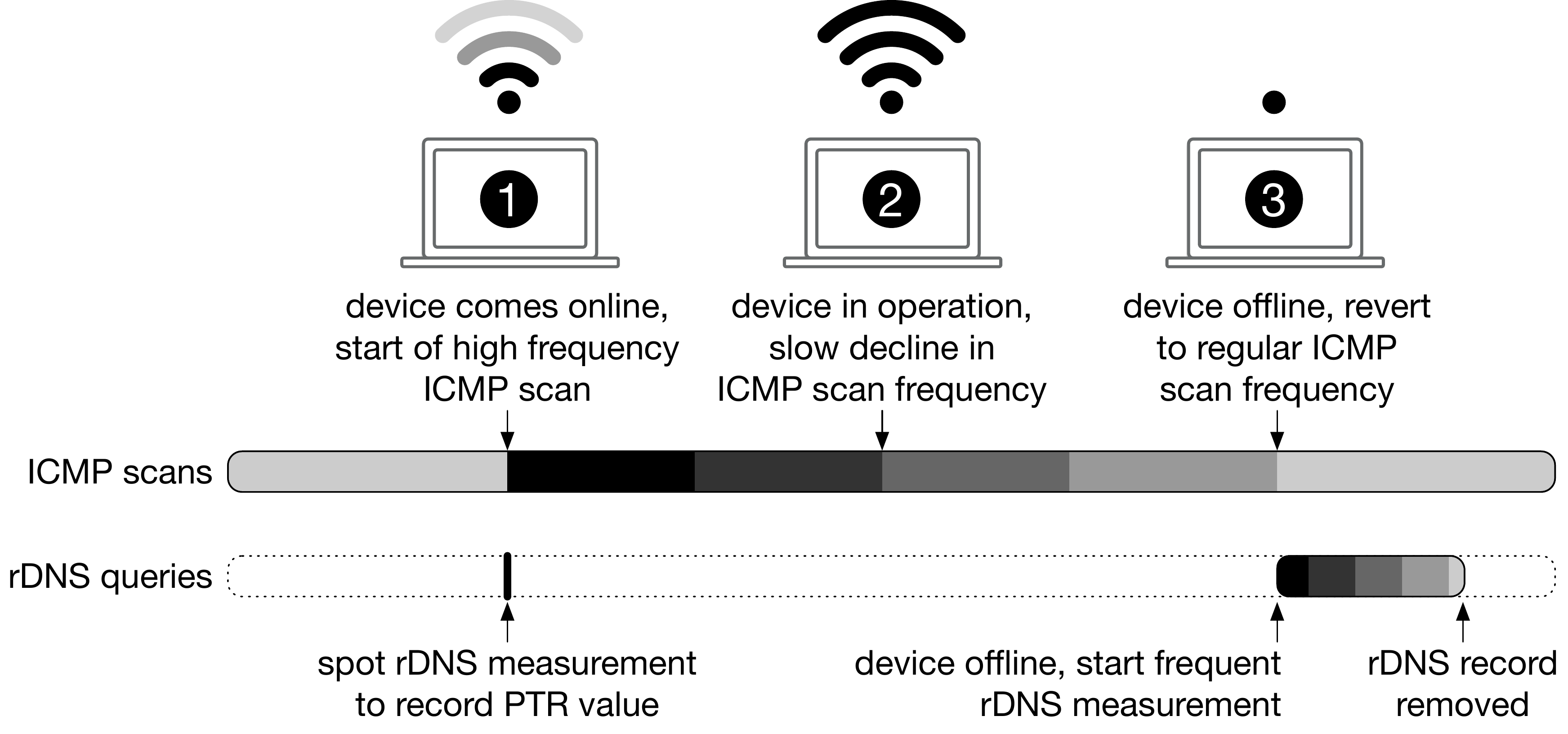}
	\caption{Graphic representation of the mechanics of the supplemental measurement. Time flows from left to right, the two bars represent when ICMP and \gls{rdns} measurements are scheduled, numbers correspond to device activity phases.}
	\label{fig:supp-measure-flow}
\end{figure}

Phases 1 and 3 can speak to the relation between the presence of a client 
and the presence of an \gls{rdns} record. To measure this, we trigger reactive measurements
when we infer, from the hourly ICMP ping scan, that a client has newly appeared on the network. 
We perform a reactive ping and trigger an \gls{rdns} query. We then continue
with pings at five-minute intervals and gradually back off. After pinging
12$\times$, which takes one hour, we reduce to ten-minute intervals during the second
hour. The full back off schema is shown in
Table~\ref{tbl:active_measurement:frequencies}.
Once we infer that the client is no longer reachable, we reactively perform
\gls{rdns} lookups for the IP address in question, following the same back off schema. Figure~\ref{fig:supp-measure-flow} shows a graphic representation of our supplemental measurement.

We use Zmap~\cite{durumeric2013} for the ICMP measurements. Zmap allows us to
easily implement rate limiting and IP address blocklisting.
The blocklisting capability is used to allow
subjects to opt-out (see Section~\ref{sec:ethical_considerations}). We have set
up our measurement infrastructure such that information about the measurement
is easily findable, and contact details are given to allow opt-out of the
measurement.

For the \gls{rdns} measurement we use custom-built software wrapping \emph{dnspython}.\footnote{\url{https://www.dnspython.org/}} 
We rate-limit requests to
authoritative name servers to reduce the impact of our measurement on the
\gls{dns} name servers as much as possible. We query the authoritative name
server for the IP address in question directly, to make sure we get a fresh answer
(i.e., not from a cache).
Both Zmap and our custom-built software write the results as CSV files to disk.
Zmap only includes hosts that were reachable in its output. 

\paragraph{Supplementary Measurement Data}
\label{ssec:active_measurement:dataprocessing}

The supplementary \gls{rdns} data may contain resolution errors, which come in
the form of \texttt{NXDOMAIN}, authoritative name servers failing to answer, or
timeouts. In our data set it is clear which are correct \texttt{PTR} responses
and which are errors.
The shortest follow-up time is five minutes. For this reason we add, next to
the original timestamp, a truncated timestamp per five minutes to the ICMP and
\gls{rdns} measurement data points. We can then merge supplementary measurement
data based on IP address and timestamp. 
We next determine the start and endpoints of client activity by relating
measurement data points. We are mostly interested in knowing what happens to
the \gls{rdns} after a client joins or leaves. However, by considering
\gls{rdns} data from around the time of the client joining, we can also verify
if \gls{rdns} state is reverted after a client has left.

We assign activity at the IP address level and give each address, start and end point
combination a group ID. This group ID allows supplementary measurement data
to be tied to specific client activity periods.
As a last step we aggregate by group ID, including the timestamps
of the last ICMP and \gls{rdns} measurements, and the first and last measured
\gls{rdns} entries.
For groups to be usable towards making inferences, each should include at
least successful ICMP probes and \gls{rdns} lookups for phases 1 and 3 (the
client joining and leaving the network). If the group's data shows that the
\gls{rdns} record is added and then removed, we can reliably infer a relation
between the \gls{rdns} record and observed client activity, which allows us to
investigate the (temporal) relation between the presence of clients on the network and the
presence of \gls{rdns} records.

\subsection{Observations of Client Activity}
\label{ssec:results:active_measurement}
\begin{table*}
    \caption{Supplemental measurement statistics.}
    \label{tbl:datasets:active_measurement}
    \begin{center}
        \resizebox{.9\textwidth}{!}{%
            \begin{tabular}{@{} l l l R{2.5cm} R{5cm} R{5cm} @{}}
\toprule
{} &  Start date &    End date &  Total \# responses &  \# of unique IP addresses observed &  \# of unique PTR records observed \\
\midrule
ICMP &  2021-10-25 &  2021-12-05 &           45,496,201 &                              80,738 &                                 - \\
rDNS &  2021-10-25 &  2021-12-05 &           11,731,348 &                              54,456 &                           180,614 \\
\bottomrule
\end{tabular}

        }
    \end{center}
\end{table*}

\begin{table}
    \caption{The 9 networks targeted for supplemental
	     measurement, along with their type, the size of the targeted address space, and number
	     of addresses that respond to ICMP probes.}
    \label{tbl:results:active_measurement:icmp}
    \begin{center}
        \resizebox{\columnwidth}{!}{%
            \begin{tabular}{@{} l R{2cm} R{2.5cm} R{2cm} @{}}
    \toprule
    {} &  Network size &  Addresses observed &  Percent observed \\
    Network name &               &                     &                   \\
    \midrule
    Academic-A   &           /16 &              31,454 &             48.0\% \\
    Academic-B   &           /16 &                   2 &              0.0\% \\
    Academic-C   &           /16 &              21,602 &             33.0\% \\
    Enterprise-A &      /17, /19 &              24,055 &             58.7\% \\
    Enterprise-B &       3 * /16 &                   0 &              0.0\% \\
    Enterprise-C &       5 * /24 &                   0 &              0.0\% \\
    ISP-A        &       3 * /22 &               1,073 &             34.9\% \\
    ISP-B        & /16, /17, /18 &                 357 &              0.3\% \\
    ISP-C        &           /16 &               1,102 &              1.7\% \\
    \bottomrule
\end{tabular}

        }
    \end{center}
\end{table}

Table~\ref{tbl:datasets:active_measurement} shows summary statistics on the
supplemental measurement, which we ran from 2021-10-27 to
2021-12-05.
Table~\ref{tbl:results:active_measurement:icmp} shows additional, per-network information
related to the supplemental measurement. The nine (anonymous) networks are shown, along
with their type, size of the targeted IP address space, and number of addresses that
respond to ICMP probes.
For two out of three \emph{enterprise} networks, we do not see responses to
ICMP pings at all. We suspect the operators of these networks block pings on ingress.
For one \emph{academic} network (\texttt{Academic-B}) only two hosts responded to ICMP
pings consistently throughout the measurement, but these particular IP addresses
do not have \texttt{PTR} records.
While we receive responses from within all three \emph{ISP} networks, the responsiveness rates
vary. We suspect that, in these networks, no blocks are imposed
by the operator, and responsiveness thus fully depends on hosts being online.

We reiterate that the networks were selected because they show strong signs
of adding dynamically assigned hostnames to \gls{rdns}. So even for the
networks (or hosts) that block ICMP probes, \gls{rdns} data \emph{can} be used to learn client device presence.
In addition, \gls{rdns} queries reveal the hostname
attached to the IP address, something ICMP probes do not provide.

\begin{figure}
    \includegraphics[width=\columnwidth]{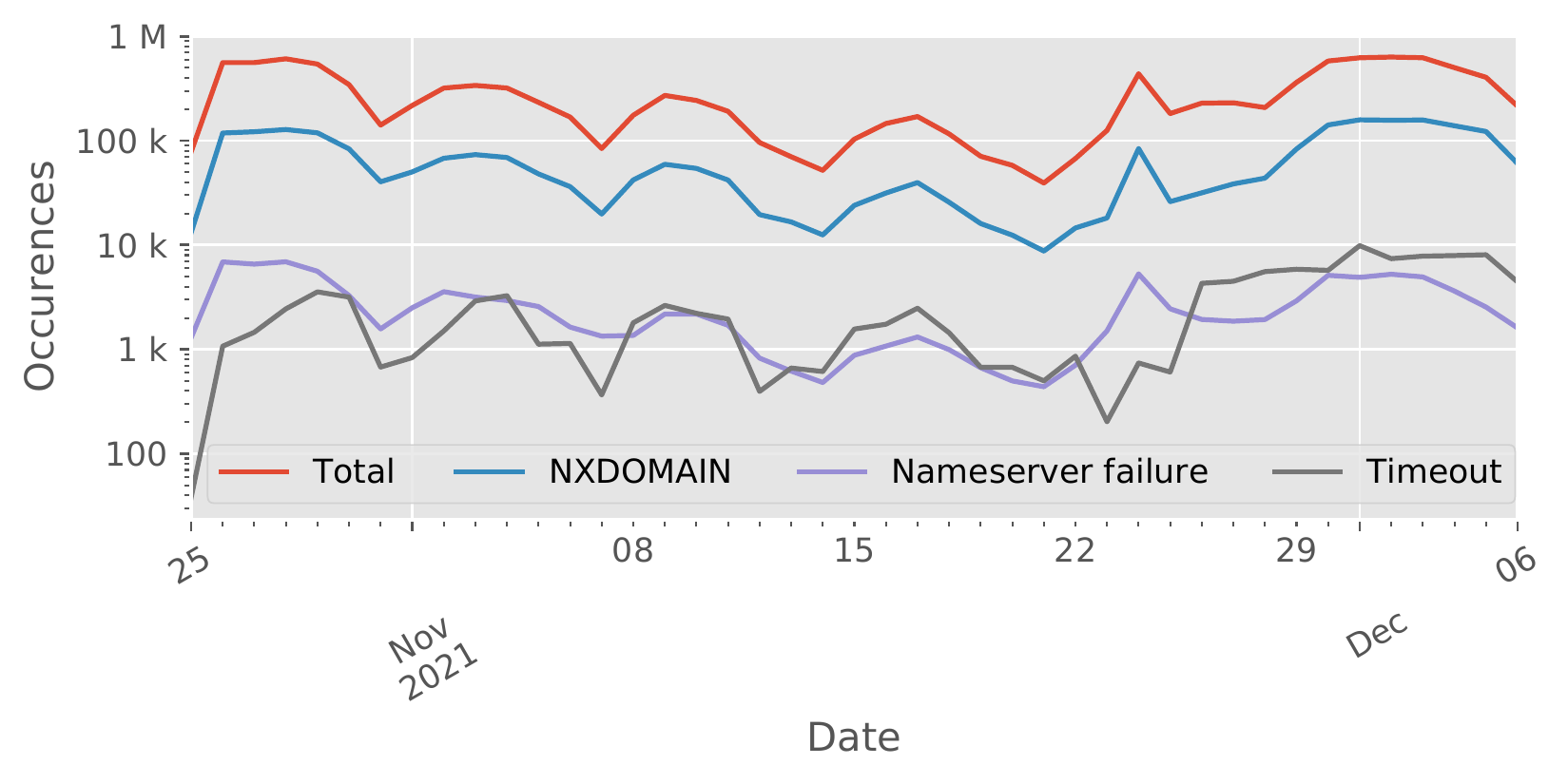}
    \caption{DNS errors observed during the supplemental measurement.}
    \label{fig:results:active_measurement:dns_errors}
\end{figure}

Our supplemental measurement resulted in errors at times. Next to the normal
responses, we saw name server failures, timeouts, and \texttt{NXDOMAIN}
responses.
Figure~\ref{fig:results:active_measurement:dns_errors} shows the number of
errors compared to the number of IP addresses seen per day (note the
logarithmic y-axis).
Fortunately, the number of errors is low relatively to the number of queries
performed. In traditional \gls{dns} sense, receiving an \texttt{NXDOMAIN}
response is seen as an error. In our case, however, this is a bit more nuanced.
Depending on the time frame, a \texttt{PTR} could be missing because it is yet
to be added to the \gls{dns} (phase 1 of client activity) or already removed
(phase 3).\footnote{Sending additional \texttt{PTR} lookups for phase 1 would
result in fewer inconclusive measurements, but the phase 3 issue cannot be
corrected for during measurement.}


\begin{table}
\caption{Breakdown of supplemental measurement results, down from all groups
	 to those enabling inferences. \reviewfix{e-4}}
\label{tbl:results:active_measurement:breakdown}
\begin{center}
    \resizebox{\columnwidth}{!}{%
	\begin{tabular}{@{} l R{2cm} R{3cm} @{}}
\toprule
{} & \#groups & Fraction of parent\\
\midrule
All groups                                           & 6,297,080 & 100.0\%\\
\quad{}Successful responses                          &   582,814 &   9.3\%\\
\quad{}\quad{}\texttt{PTR} reverted                  &   581,923 &  99.9\%\\
\quad{}\quad{}\quad{}Reliable timing alignment       &   419,453 &  72.1\%\\
\bottomrule
\end{tabular}

    }
\end{center}
\end{table}

As explained in our methodology
(Section~\ref{ssec:active_measurement:dataprocessing}), we group supplementary
measurement data points.
Table~\ref{tbl:results:active_measurement:breakdown} shows a
breakdown of the groups in the supplementary data, starting with all measurement groups, down to those that can be used to make reliable inferences.
\reviewfix{e-4}The first group subcategory, \emph{successful responses}, ensures the group
has successful ICMP probes and \gls{rdns} lookups for the client joining and
leaving the network (i.e., no timeouts or errors).
The next subcategory, \texttt{PTR} \emph{reverted}, involves groups in which we
observe that the \texttt{PTR} is changed and reverted during and after the client's
inferred presence.
Of this subcategory, in about 1 out of 4 cases, timing mechanics of the
ICMP probes, which cannot be accounted for at run-time without compromising the back off mechanism,
make the results less reliable.
After filtering these out, we are left with 419,453 usable groups.


\paragraph{Validating Timing Aspects}

\begin{figure}
    \begin{subfigure}[b]{\columnwidth}
        \includegraphics[width=.9\columnwidth]{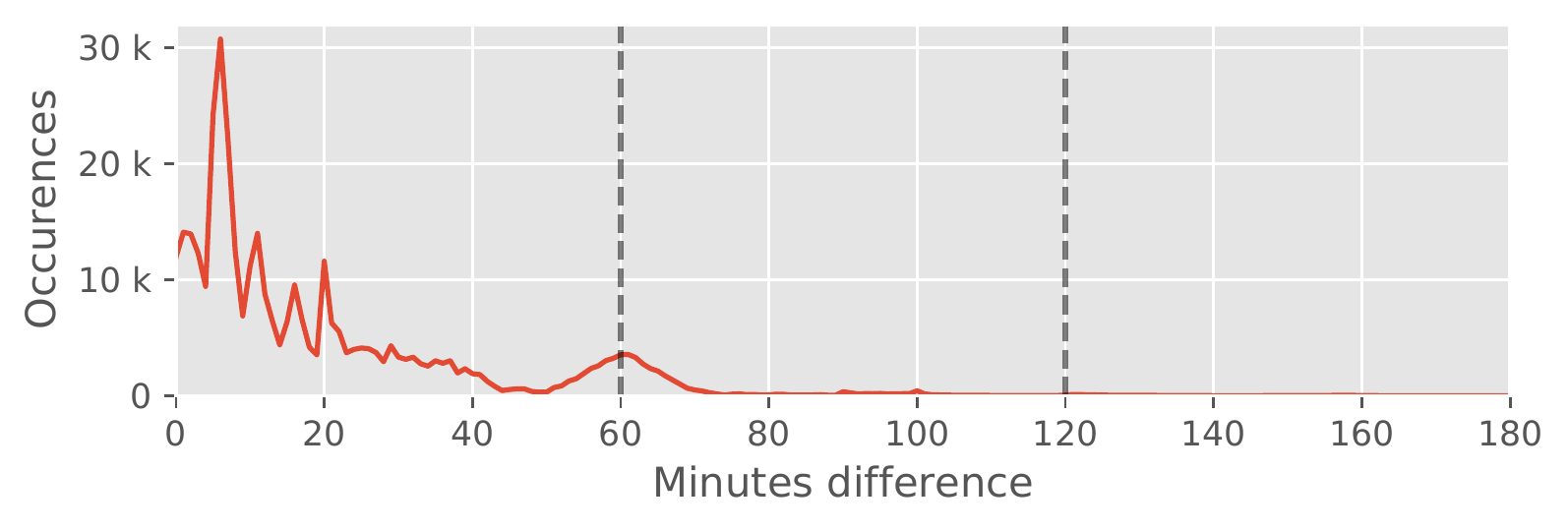}
         \caption{Absolute numbers of groups for a given (x-axis) minutes
                  difference. First three hours are
                  shown.}
          \label{fig:results:active_measurement:difference:occurences}
    \end{subfigure}
    \begin{subfigure}[b]{\columnwidth}
        \includegraphics[width=.9\columnwidth]{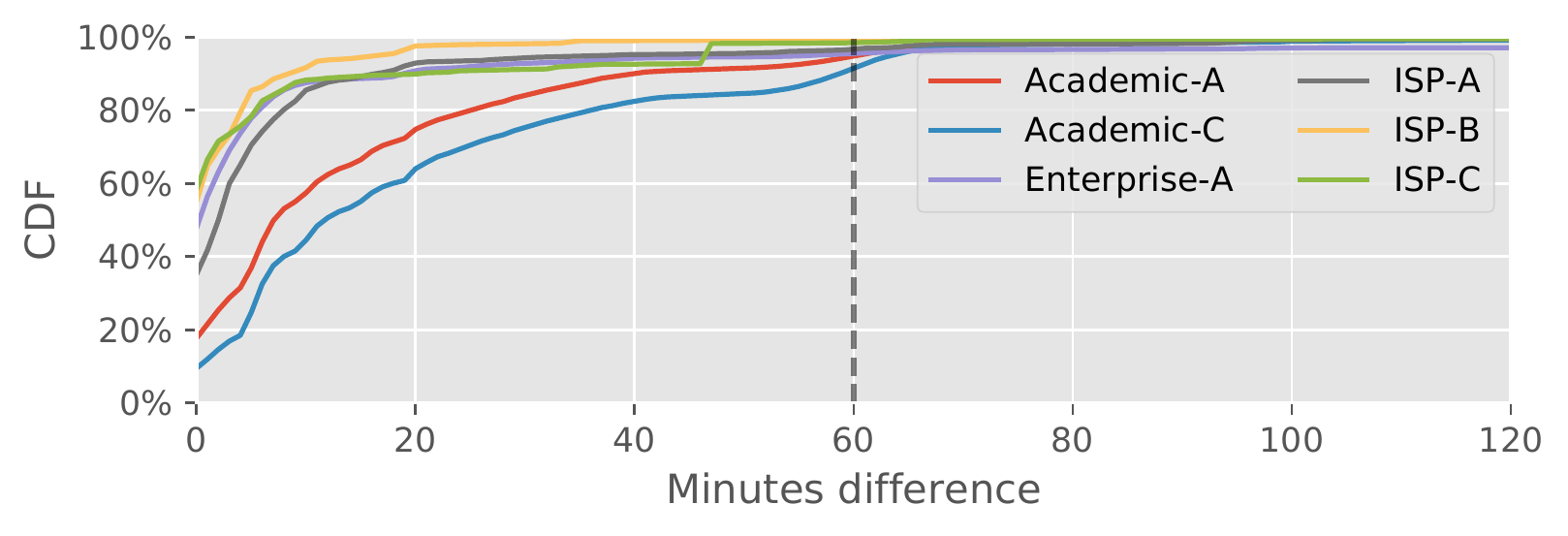}
        \caption{CDF of the differences per network shown for the first two hours.}
	 \label{fig:results:active_measurement:difference:cdf_networks}
    \end{subfigure}
    \caption{Difference in minutes between last ICMP sample and rDNS sample per measurement group.}\label{fig:results:active_measurement:difference}
\end{figure}

We can now shed light on the temporal relation between the presence of clients
on the network and the presence of \gls{rdns} records.
Figure~\ref{fig:results:active_measurement:difference:occurences} shows the number
of groups offset against the time between a client leaving
the network and \texttt{PTR} removal.
The peaks around multiples of an hour suggest that \texttt{PTR} records are
removed due to the expiration of a DHCP lease, which is often set to an hour
for a fast turn-over rate.
The peak close to the five minutes mark can be explained by clients cleanly
leaving a network (i.e., by sending the optional \gls{dhcp} release message
upon leaving the network). If the \gls{dhcp} server or \gls{ipam} system removes
the \texttt{PTR}, we would see this five minutes later as the (respective) probes
are sent in five-minute intervals.

Figure~\ref{fig:results:active_measurement:difference:cdf_networks} breaks down
timing information for the individual networks targeted for supplemental
measurement. The networks \emph{Enterprise-B} and \emph{Enterprise-C} are not
shown, because no ICMP ping responses were received from these networks. \emph{Academic-B} is not shown because the two hosts responding to ICMP
did not have a corresponding \gls{rdns} entry.
The CDFs show that in about 9 of 10 cases, the \gls{rdns} entries reverted within 60
minutes of a client leaving the network. We already demonstrated that the presence of
client devices on the network can be learned from \gls{rdns}, which
anyone can query. The fact that records do not linger long in most cases escalates the privacy risk: the timing
enables observation of network dynamics.
The differences in lingering between \emph{Academic-A} and \emph{Academic-B},
as apparent in Figure~\ref{fig:results:active_measurement:difference:cdf_networks},
can be explained by a longer \gls{dhcp} lease time. If these networks predominantly
update \gls{rdns} in response to leases expiring, rather than to \gls{dhcp}
release messages, the \gls{rdns} records linger longer.

$ $\\
\textbf{Key takeaways:} \emph{In networks that expose the presence of dynamic
clients, there is a strong link between the existence of a \texttt{PTR} record
for an IP address and the presence of a client device, which has been assigned that address. In 9 out of 10
cases, \texttt{PTR} records linger for 60 minutes or less after client disappearance.}


\section{Case studies}
\label{sec:case_studies}

Now that we have demonstrated that \gls{rdns} entries can be used to infer the
presence of client devices on networks, we present a number of case studies.
These case studies serve to look at some of the consequences and further
substantiate the privacy risk.

\subsection{Life of Brian(s)}
\label{sec:case_studies:brians}

\begin{figure}
    \begin{center}
        \resizebox{\columnwidth}{!}{%
            \includegraphics[width=\textwidth]{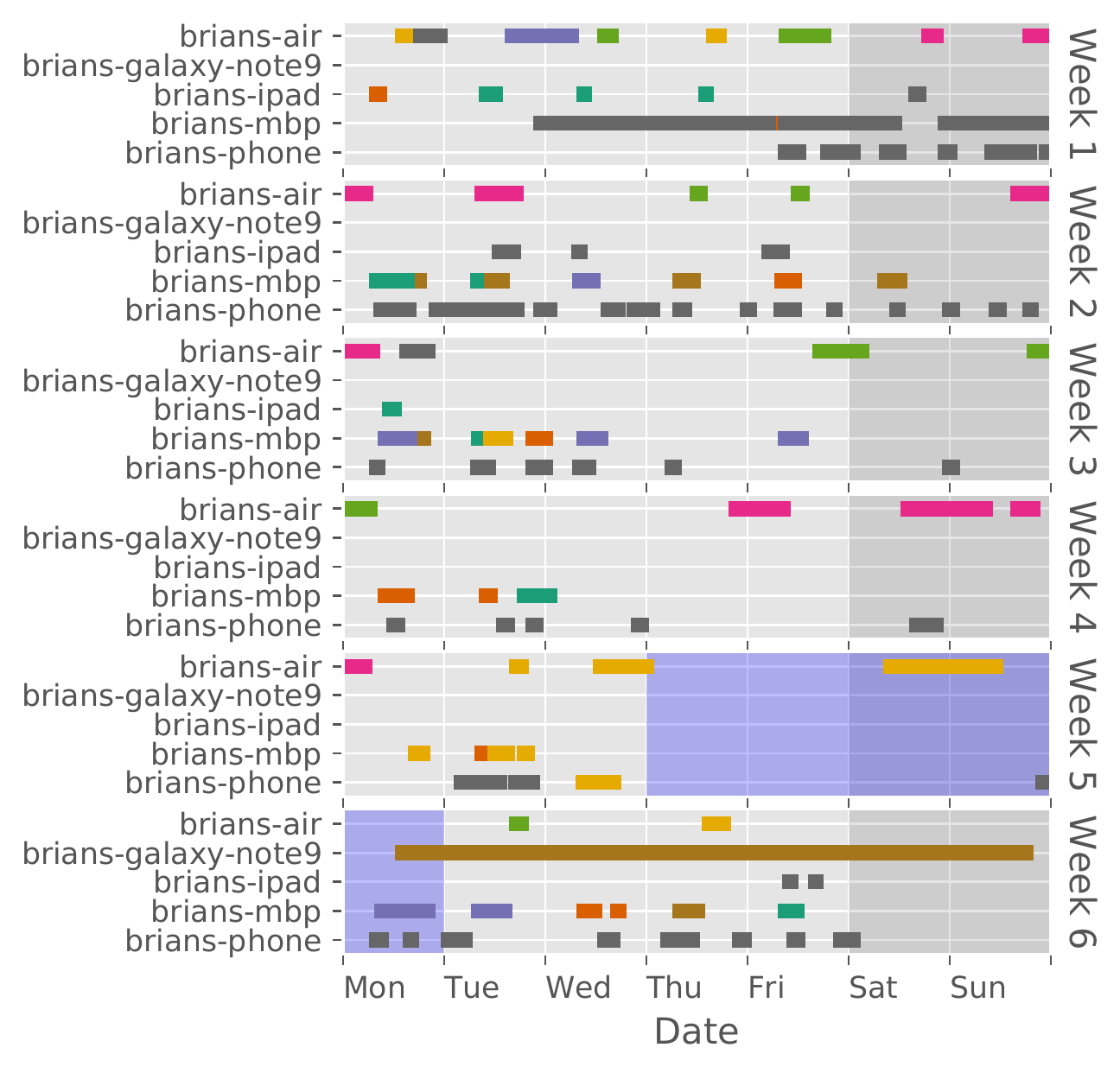}
        }
    \end{center}
    \caption{Six weeks in the Life of Brian(s). Weekends (gray) and Thanksgiving weekend (purple) are highlighted through shading. IP addresses are encoded with colored bars.}\label{fig:casestudies:brian:life}
\end{figure}

To demonstrate the severity of the privacy risk, we use \gls{rdns} data to follow
persons named \emph{Brian} over time. For this we assume that the
\emph{given name} in the hostname reflects the name of a network client's owner.
We use our supplementary \gls{rdns} measurement data for this case study.  It
is important to note that while we reactively looked up \texttt{PTR} records
with ICMP pings as the trigger point during supplemental measurement, anyone
with the capability to do frequent \texttt{PTR} lookups can capture the
same patterns that we discuss in this case study (i.e., no ICMP required).

We use the data of the network \emph{Academic-A}, which is an academic
network in the US with campus housing.
Figure~\ref{fig:casestudies:brian:life} shows six weeks of client hostnames
containing the \emph{given name} \texttt{Brian} on \emph{Academic-A}. We have
color-coded IP addresses in the figure. Times in the figure are in the local timezone.
Our intuition is that these hostnames are not related
to a single \emph{Brian}, but rather two or maybe three. The use of a private
and work phone is not uncommon, but multiple laptops in active use arguably is.
The clients \texttt{brians-air}, \texttt{brians-mbp} and \texttt{brians-phone} show regular
patterns. Especially \texttt{brians-mbp} in week two shows very regular activity:
a couple of hours around noon, every day.

We chose \emph{Academic-A} and these six weeks because of the Thanksgiving weekend
(the weekend of the fifth week). Thanksgiving is a US holiday in which many
students go home to be with their families. Thanksgiving is always on a
Thursday. In 2021, it fell on the 25\textsuperscript{th} of November.
The Friday and Monday after Thanksgiving are known as \emph{Black Friday} and
\emph{Cyber Monday}, and many stores promote sales around these days, typically
on electronics.
In our results, \texttt{brians-phone} and \texttt{brians-mpb}
seem to leave the network around Thursday. Striking is that
\texttt{brians-galaxy-note9} appears in the afternoon on Cyber Monday. We have
not observed this hostname before this time. 
We speculate that a Brian may have bought
a Samsung Galaxy Note 9 during the Black Friday or Cyber Monday sales.

This case study shows that \gls{rdns} data provides insights into the behavior
of clients to which hostnames are dynamically assigned.  Since the hostnames
contain given names, this may even be tied to specific individuals.
If one knows or were able to infer how addresses are assigned to, e.g.,
specific buildings~\cite{zhang2021inferring} on campus, one could track, from
virtually anywhere on the Internet, a \emph{Brian} around campus as he goes
from lecture to lecture.

\subsection{Working from Home}
\label{sec:case_studies:wfh}

For the second case study, we move away from tracking specific
(\emph{Brian}-owned) clients over time and focus on investigating the overall dynamics
of select networks instead.
We use the OpenINTEL-provided \gls{rdns} data for this case study, to
demonstrate that daily \texttt{PTR} measurements already offer insight into
network dynamics.
We select all three \emph{academic} networks for this case study, as well as
\emph{enterprises} B and C, as these five networks show a specific change in
user behavior that we wish to highlight in this case study.

\begin{figure}
    \resizebox{\columnwidth}{!}{%
        \includegraphics[width=\columnwidth]{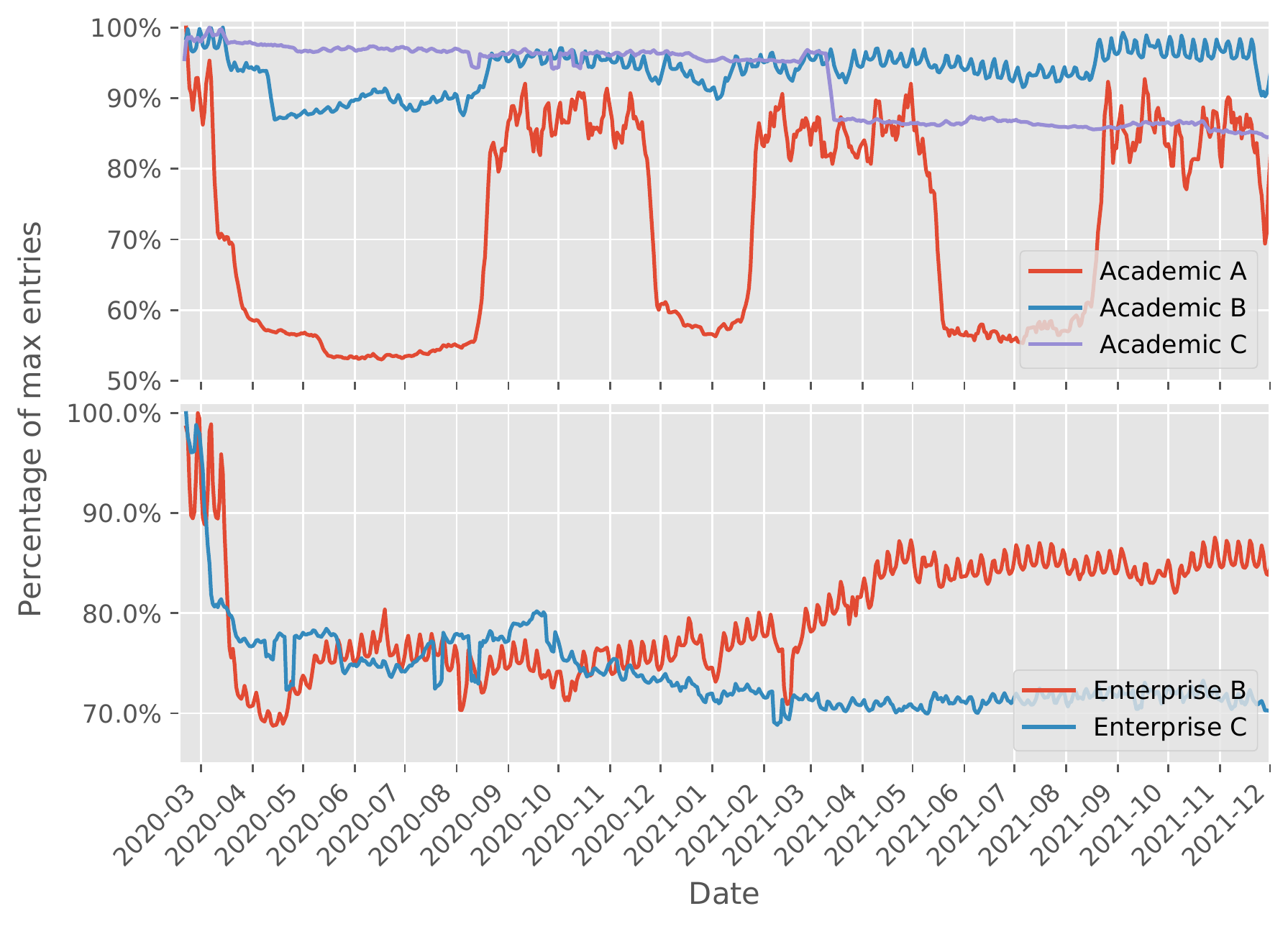}
    }
    \caption{A longitudinal breakdown of reverse \gls{dns} entry presence for select networks, starting near the beginning of the COVID-19 outbreak. The
	     percentage shown are of the number of entries relative to the maximum observed
	     number of entries in each network.}
    \label{fig:casestudies:wfh:wfh}
\end{figure}

For each of the selected networks, we calculate the total number of \texttt{PTR} records on
any given day (i.e., we do not require given names to be present, etc.).
Our aim is to see if there is a correlation between the presence of \gls{rdns}
entries and lockdown regulations due to the COVID-19 pandemic.
We expect enterprise networks to experience a drop in daily entries as employees were
required to work from home. For academic networks this may be a bit more
nuanced. Education buildings may see fewer clients, but with on-campus student
housing, the client concentration may shift without it necessarily being
visible in the total number of \gls{rdns} entries.

Figure~\ref{fig:casestudies:wfh:wfh} shows for each network the percentage of
\gls{rdns} entries relative to the maximum number observed, over 2020 and 2021. The three academic networks
are shown at the top. The selected enterprise networks at the bottom.
We compare the public COVID-19 related news reports of \emph{Academic-A} (shown
in red in the figure) with the presence of \gls{rdns} entries. Upon making this
comparison, we see a clear connection between the two.\footnote{We do not link to these reports
to protect the identity of network \emph{Academic-A}.} 
For the times at which a moderate or high risk was reported to students and
staff, sharp decreases in daily \gls{rdns} entries records are visible. After
reports of a low risk of COVID-19 prevalence on campus, a sharp increase in
network client device presence is visible.

For \emph{Academic-B} we observe a marked reduction in \gls{rdns} entries during
the first period of COVID-19 lockdowns, after which the number
goes back up to about 95\% of what we observed before the start of the pandemic.
By September~2021, the level returns to that of before the pandemic, with
the dip at the end corresponding to the Christmas holiday break.

The networks of \emph{Enterprise-B} and \emph{Enterprise-C} show significant
decreases in \gls{rdns} entries in March and April of 2021. It stands to reason
that these decreases are related to COVID-19 measures.
\emph{Enterprise-B} shows a partial recovery in the number of entries around
May of 2021, which could be a sign of loosened restrictions, either
by the government or by the employer.

\begin{figure}
    \resizebox{\columnwidth}{!}{%
        \includegraphics[width=\columnwidth]{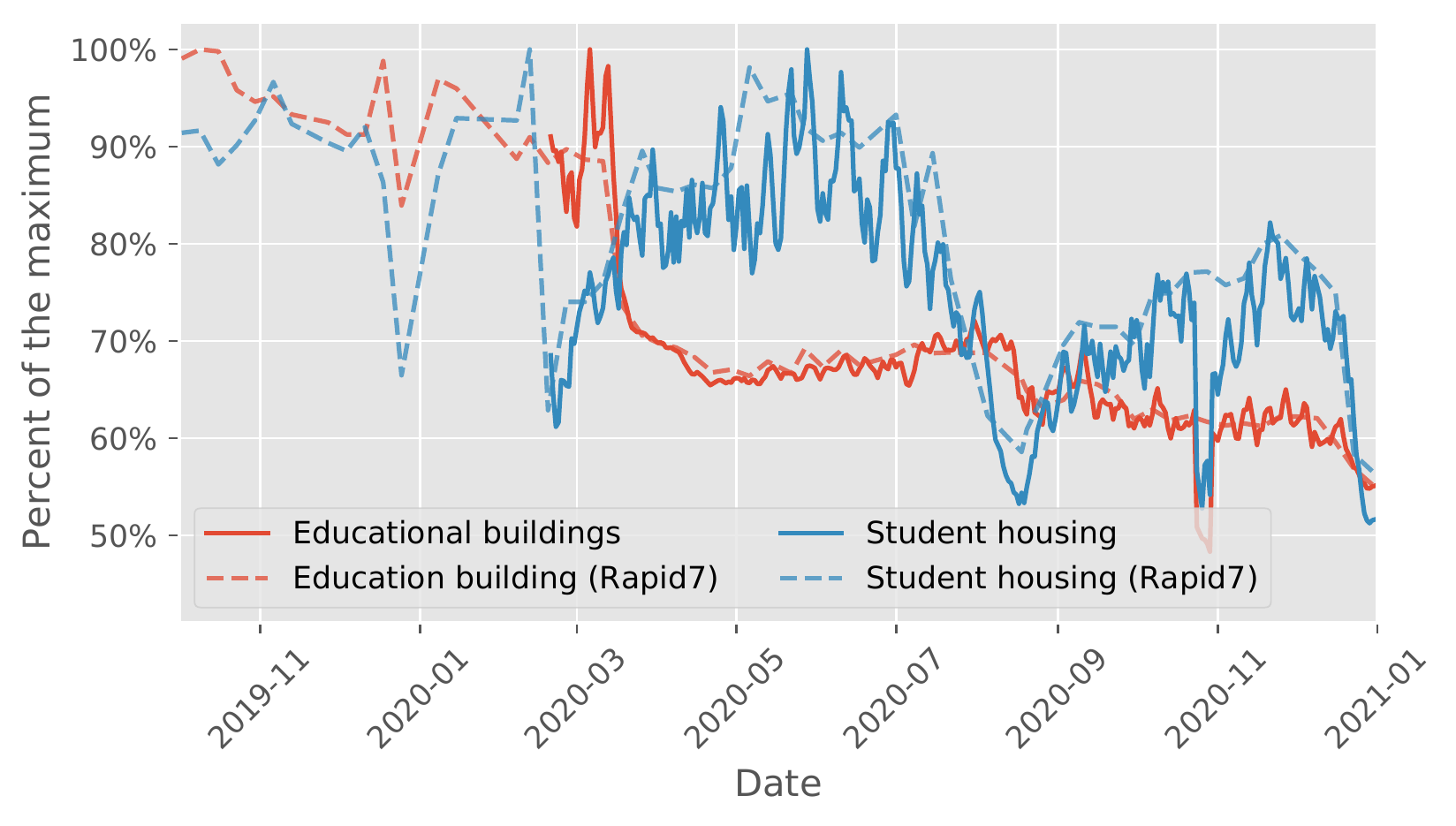}
    }
    \caption{Zoomed in version of \texttt{Academic-C}, starting late 2019. Dashed
             lines are based on Rapid7 data. Solid lines are based on OpenINTEL data.
             Rapid7 provides weekly snapshots of rDNS state. OpenINTEL started daily
             rDNS measurements in February 2020.}
    \label{fig:casestudies:wfh:academic_c}
\end{figure}

In Figure~\ref{fig:casestudies:wfh:academic_c} we look at network
\emph{Academic-C} in more detail. As this is the home institution of
the authors, we know which IP subnets are used for on-campus student housing
and educational buildings, and when buildings were closed.
In this graph we show both Rapid7 and OpenINTEL data. The COVID-19 lockdown
measures were introduced shortly after OpenINTEL started its
\gls{rdns} measurements (2020-02-17). We use Rapid7 data, which has weekly granularity, to
extend visibility into the early months of 2020.

In March a crossover between \texttt{PTR} records for educational buildings and
student housing is clearly visible, signifying that: employees are working from home, educational
buildings are empty, and students study from their on campus residences.
The decreases at the end of October for both the educational
buildings and student housing corresponds with the fall holiday week. A similar
drop is visible at the end of the year for the 2020 Christmas break.

In absolute numbers (not in figure), the reverses for the educational buildings
remain much higher than for student housing, which can be explained by having
more address space assigned to educational buildings, with more static hosts
online.
The Rapid7 curves in Figure~\ref{fig:casestudies:wfh:academic_c} (dashed)
largely overlay and confirm the observations we make from OpenINTEL data.
Additionally, given that Rapid7 data extends visibility into 2019, we can see that the
number of network clients before the crossover is relatively
stable. The 2019 Christmas break is also visible in Rapid7 data, as well as a drop towards the
end of February 2020 that likely relates to Carnaval celebrations.\footnote{A local Catholic holiday.}

While our case study into compliance with work-from-home measures is
relatively harmless, it does show the extent to which even \gls{rdns} measurement data of daily granularity
can be used to learn network dynamics, possibly for nefarious purposes.
\reviewfix{c-2} Our approach to observing work-from-home patterns using
\gls{rdns} data adds adds to existing efforts in the literature towards this end (e.g., observing shifts in traffic volumes).

\subsection{When to stage a heist?}
\label{sec:case_studies:heist}

Suppose that you want to stage a heist. There is something valuable in a
building and you want to steal it while the least amount of people are around.
Ideally, from the robber's perspective, they are able to determine the point in
time at which the fewest dynamic clients are around.
This evidently requires high-frequency \gls{rdns} measurements. For ethical
reasons, we have not instrumented such a measurement. For this case study, we
rely on data from our supplemental measurement instead.

We consider one week of supplementary data for the network \emph{Academic-A}.
This network responds to ICMP pings, which we use to support our findings. We
stress, again, that ICMP responsiveness is not required. Even networks that
block ICMP may be observed via \gls{rdns}, as our second case study shows
(recall from Section~\ref{ssec:results:active_measurement} that networks
\emph{Enterprise-B} and \emph{Enterprise-C} do not respond to ICMP probes).

\begin{figure}[t]
    \resizebox{\columnwidth}{!}{%
        \includegraphics[width=\columnwidth]{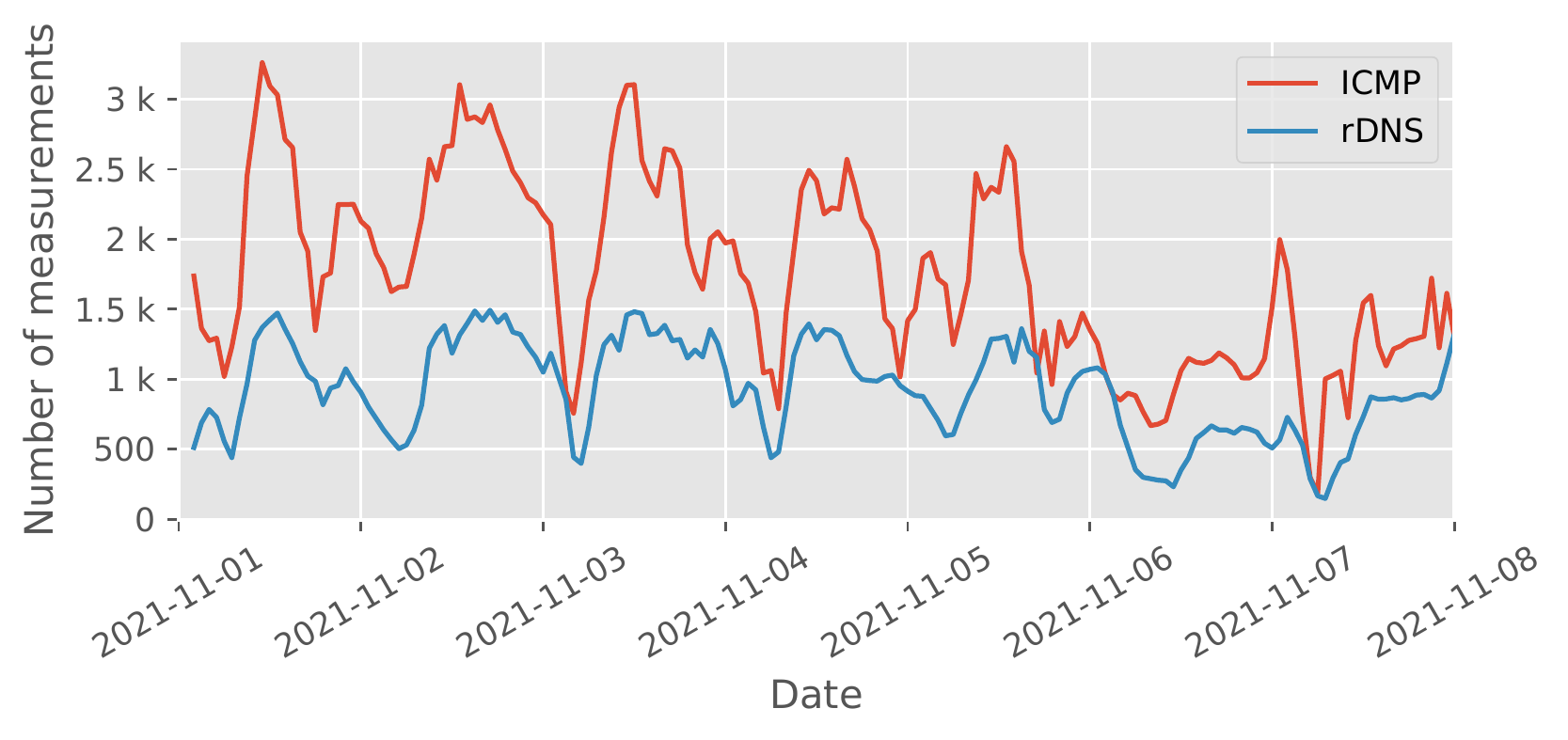}
    }
    \caption{One week of measurements from \emph{Academic-A} to demonstrate
             when one might stage a heist.}
    \label{fig:casestudies:burglary}
\end{figure}

Figure~\ref{fig:casestudies:burglary} shows the number of active clients
inferred in the network \emph{Academic-A} between 2021-11-01 and 2021-11-07, both
for \gls{rdns} lookups (blue) and ICMP probes (red). The diurnal pattern of the network
is visible, with most activity during the day and into the evening, while the
least activity is at night and early in the morning.
The \gls{rdns} measurements (blue) give a rough indication of the best
time for the heist. As an example, on weekdays the data hint at approximately
6AM as a good time.
We also show activity based on ICMP responses (red) for comparison and to
support our findings. The ICMP results for the most agree. For networks that do
not block ICMP, the robber could of course also use ICMP probes. 
In absolute numbers, the \gls{rdns} lookups pan out lower than the
number of ICMP probes, which is due to the reactive nature of the \gls{rdns}
measurement.

This case study shows the feasibility of one example of how outside
observations of dynamically assigned hostnames can be used for nefarious
purposes.
\reviewfix{c-3} These observations can help a potential attacker to learn
working patterns without being physically present at the location.
Our supplemental measurement is reactive and does not try to establish the
number of clients on a network at any given time. A targeted measurement at a
higher frequency would likely give better results. We leave a study to confirm
this as future work, as this is out of scope for this paper.


\section{Discussion}
\label{sec:discuss}
Our findings are disconcerting. While existing literature has shown that meaningful
information can be extracted from hostnames primarily without considering continual changes to reverse \gls{dns} records, we reveal
that observing automated changes to \gls{rdns} can provide
insights into client presence and network dynamics.
\reviewfix{e-sum-5|b-3}
The publicness of \gls{rdns} severely increases this risk, enabling anyone on the 
Internet to observe automated changes. An adversary with measurement
capability and knowledge about a potential target can gain valuable insights
following an approach similar to ours.
We keep the invasiveness of our case studies in check and tailored our approach,
but given recent findings that hostnames can encode building
locations~\cite{zhang2021inferring}, it appears feasible that for some networks,
\gls{rdns} data can be used to geotemporally track users at the building level.

An arguably sensible mechanism to limit the tracking of network client
devices by outsiders is blocking ICMP pings at network ingress. Two of the
networks we used for validation do not respond to ICMP pings. At the same time,
the records these networks dynamically add to the global \gls{dns}, as well as the
time these records linger after clients have left, allow anyone who frequently
queries for \gls{rdns} records to observe the presence of clients in these networks.

A notion that other works that study hostnames have in common is that 
meaningful information is encoded in hostnames on purpose, especially for router-level entries.  Our results
substantiate that the interplay between \gls{dhcp} and \gls{dns} can
inadvertently provide anyone with \gls{dns} lookup capability insights into
end-user client identifiers.
Our results also reveal a more severe problem: privacy-sensitive information such as
device owner names appear in the \emph{global} \gls{dns}. While our validation
covers nine networks, this problem is likely not limited to these networks.
Our results thus confirm a risk outlined in RFC~7844~\cite{RFC7844}: \gls{dhcp}
clients send revealing information in optional parameters. Based on our observations of terms
commonly co-appearing with \emph{given names} (e.g., \texttt{brians-ipad}
and \texttt{brians-galaxy-note9}), we suspect that client implementations on
various makes and models of phones and computers send \emph{device names}
to the \gls{dhcp} server.
\reviewfix{e-sum-2} While our choice to match against popular names given to US
newborns creates bias towards these names, we accept this bias as we set out to
substantiate the privacy risk rather than to exhaustively identify all
\gls{rdns} records that contain privacy leaks.

\paragraph{Steps towards Mitigating the Problem}
After raising these issues we would like to start the discussion on how to
solve the problem.
Evidently, the interplay between \gls{dhcp} and \gls{dns} and the extent to
which configuration and protocols permit client identifiers to flow from one protocol to
another is at the core of this problem. While we have not investigated
this extensively, we identified a number of \gls{ipam} softwares that make it
easy to automate \gls{dns} changes. For example:
Bluecat,\footnote{\url{https://bluecatnetworks.com/}} Efficient
IP,\footnote{\url{https://www.efficientip.com/}} Infoblox,\footnote{\url{https://www.infoblox.com/}} Men \&
Mice,\footnote{\url{https://menandmice.com/}} and
Solarwinds.\footnote{\url{https://www.solarwinds.com/}}
It is unclear to us if and which \gls{dhcp} servers or \gls{ipam} systems
come with default settings that carry over client identifiers to the
global \gls{dns}.
We would argue that it is rarely a good idea to indiscriminately carry over \gls{dhcp} client-provided information such as \emph{device names} to
publicly accessible \texttt{PTR} records. Using some sort of hash seems prudent,
although this may make hostnames less sensible.
While we have not thoroughly investigated reasons for device manufacturers to
send \emph{device names} to \gls{dhcp} servers, we know that for Bluetooth
and Wi-Fi Direct pairing, sharing such information helps identify the device
in question.
The \gls{dhcp} \texttt{Host Name} option is commonly used for identification
and to update the address of the host in \emph{local} name services (see
Section~\ref{ssec:related_work:background}). The \texttt{Client
FQDN}~\cite{RFC4702} in turn can instrument \emph{global} changes, if the client so
desires. An open question is which option devices send identifying info
in, why, and whether or not this is used as intended.

A large part of the problem is the practice of dynamically adding \texttt{PTR}
records. While unique identifiers in \texttt{PTR} records enable
tracking of specific clients, even record presence in itself provides
insights into network dynamics, which combined with other information (e.g.,
knowledge about building-level IP subnet assignments) stands to reveal a lot.
This should be better understood by network operators.
\reviewfix{e-sum-3-1}
Our advice to network operators to reduce the harm of this problem is to block
the propagation of \texttt{Host Name} information from \gls{dhcp} to \gls{dns}.
This can be done by reviewing and adapting the configuration of the internal networks.
\reviewfix{e-sum-5-b|d1}
IPv6 configuration also requires attention. In our study, we mainly focus on
IPv4 addresses due to the complexity of efficiently scanning the IPv6 address
space at scale.  However, an attacker willing to infer information on a
specific target can leverage previous studies (e.g., Borgolte et
al.~\cite{Borgolte2018}) to perform a targeted scan. In addition, when focusing
on IPv6, attention should be put on investigating the interaction between
domain names and IPv6 addresses when SLAAC and stateless DHCPv6 are in use
\cite{rfc4941}, since this can lead to even more fine-grained tracking of
a specific host, thus increasing privacy risks.


\section{Ethical Considerations}
\label{sec:ethical_considerations}

We follow best practices with regards to Internet measurements during our study. Where possible, we rely on existing data sources (Rapid7, OpenINTEL) rather than conducting our own Internet-wide scans. For the supplemental measurement we use to augment the existing data sets, we ensure that we rate limit our scans, we only target address space in networks that we know to be used for dynamic address allocation, we set up a web page on the scanning host that clearly explains the purpose of our study and we act immediately on requests from operators to opt out of our measurements.

\paragraph{IRB Approval}
In addition to following the best practices outlined above, our study was reviewed by our institution's ethics board prior to the start of the study. We brought up two concerns in our request to the IRB. First, even though the data we process is publicly accessible and anyone can query it, the purpose of our study and the way we perform targeted analysis raises obvious privacy concerns. Given that our goal precisely is to demonstrate these concerns --- that dynamic updates to \gls{rdns} data raises privacy issues --- this is unsurprising. In order to mitigate this concern, we store our analysis results in compliance with the EU's General Data Protection Regulation and delete data after the research concludes. We note that while this removes the immediate privacy concern of the analysis, in which we pinpoint individual users, the actual privacy threat remains, since anyone can still query the data and reproduce our results. To further mitigate this concern, we do not disclose the names of organisations whose networks we selected for further study, and we take care to report on users in aggregate only. Finally, when zooming in on individual given names, we deliberately pick a very common name.

Second, we sought express permission from our IRB for our supplemental measurement, as this measurement further aggravates the privacy risk by obtaining more timely information about the presence of devices and users on targeted networks. The IRB approved the supplemental measurement, {\it under the express condition that we minimise the number of networks to which we apply this measurement}. We detail how we select a minimal set of networks to cover with the supplemental measurement in Section~\ref{ssec:methodology:active_measurement}. In the interest of reproducibility, we retain the data from our supplemental measurement in encrypted form on our institution's servers.

\reviewfix{e-5}
Finally, we limit our analysis of specific use-cases (i.e., Life of Brian) to a
single name to minimize the possible privacy leak.  These analyses can be
conducted on any name. However, exploring this goes beyond the scope of our IRB
approval.

An approval record from our institution's ethics board is available under registration number \textit{RP 2021-202}.

\section{Conclusion and Future work}
\label{sec:conclusion}

In this paper, we performed a first of its kind study into the privacy implications of
combining \gls{dhcp} exchanges with dynamic updates to the global \gls{dns}.
Our findings do not only substantiate existing concerns that unique \gls{dhcp}
client identifiers can carry over to the \gls{dns}, but also reveal that
reverse records are based on privacy-sensitive information such as client
device owner names, and their makes and models. \reviewfix{e-sum-5} This implies that individuals, and the presence of devices likely belonging to such individuals, are at the risk of being tracked.
We analyzed the temporal relation between the presence of \gls{rdns} records
and the presence of client devices in a network and showed that -- for a selection
of academic, enterprise and ISP networks alike -- records tend to linger for at most
one hour after clients leave the network.
Via three case studies, we demonstrate that virtually anyone on the Internet
can infer and track the presence of specific clients and observe network
dynamics via reverse \gls{dns}, even with other mechanisms to limit tracking in place.
Finally, we began a discussion about the finer details of possible causes and
identified ways to start mitigating this problem.

\paragraph{Future work}

To aid mitigation efforts, there is dire need for an investigation into how \gls{dhcp} server and
\gls{ipam} software carry over privacy-sensitive client identifiers to the
global \gls{dns}. This is especially problematic if carry-over results from default
settings or settings not well understood by network operators.
One could investigate from which \gls{dhcp} option the information is taken
(e.g., \texttt{Host Name} or \texttt{Client FQDN}), if this goes against the
intended use of the option, and whether or not client-signalled \emph{desires}
for servers not to update \gls{dns} records are followed.
On the side of \gls{dhcp} client implementations, one could investigate which
\gls{dhcp} options are filled with \emph{device names}, whether or not this is
necessary, and which steps vendors can take to (partly) mitigate the problem.
Identifying exposing implementations and client devices could be done in a lab
or by using \gls{dhcp} data collected inside networks.

In this paper we did not aim to exhaustively identify exposed networks with the
highest possible accuracy. Rather, we took the first steps towards studying
the problem and used \emph{given name} matching as a starting point to drill down in 
\texttt{PTR} records.  Future work could go in the direction of investigating the full
extent of the problem, by applying techniques from related work to find patterns in hostnames.
Another angle could be to consider \emph{forward} \gls{dns} data, which can also
be dynamically updated by \gls{dhcp} servers.
Future efforts, provided that they are ethically feasible, could also focus on
the potential for fine-grained geotemporal user tracking. We used {\it a posteriori}
knowledge on IP subnet allocations in one case study, but related work has
established that access network topology can be inferred from
hostnames~\cite{zhang2021inferring}. We imagine that combining such
information with client presence inferences can have far-reaching privacy
implications.
Finally, our findings support that \gls{dhcp} release messages have an effect
on the time \gls{rdns} records linger. Future work could study the behavior of
clients in this respect: do clients that \emph{can} send releases actually
\emph{do} so and is, instead, not doing so a possible defense mechanism?

    \bibliographystyle{ACM-Reference-Format}
    \bibliography{common/library}
    
\end{document}